\documentclass[aip, jcp, amsmath, amssymb,reprint]{revtex4-1}

\usepackage[version=3]{mhchem}
\usepackage{graphicx}
\usepackage{dcolumn}
\usepackage{bm}
\usepackage{amssymb}
\usepackage{amsmath,amsfonts,epsfig}
\usepackage[english]{babel}
\usepackage{url}
\usepackage{subfigure}
\usepackage{color}
\usepackage[usenames,dvipsnames]{xcolor}
\usepackage{multirow}
\usepackage{enumerate}
\usepackage{gensymb}

\newcommand{\cref}[1]{(\ref{#1})}

\newcommand{\rcite}[1]{[\onlinecite{#1}]}
\newcommand{\blackurl}[1]{\url{#1}}

\begin{document}
\author{Anjan P. Gantapara}
\email{A.P.Gantapara@uu.nl}
\affiliation{Soft Condensed Matter, Debye Institute for Nanomaterials Science, Utrecht University, Princetonplein 5, 3584 CC Utrecht, The Netherlands}

\author{Joost de Graaf}
\affiliation{Institute for Computational Physics, Universit\"{a}t Stuttgart, Allmandring 3, 70569 Stuttgart, Germany}
%\affiliation{Soft Condensed Matter, Debye Institute for Nanomaterials Science, Utrecht University, Princetonplein 5, 3584 CC Utrecht, The Netherlands}

\author{Ren\'e  van Roij}
\affiliation{Institute for Theoretical Physics, Utrecht University, Leuvenlaan 4, 3584 CE Utrecht, The Netherlands}

\author{Marjolein Dijkstra}
\email{M.Dijkstra1@uu.nl}
\affiliation{Soft Condensed Matter, Debye Institute for Nanomaterials Science, Utrecht University, Princetonplein 5, 3584 CC Utrecht, The Netherlands}

\title{Phase Behavior of a Family of Truncated Hard Cubes}

\begin{abstract}
In continuation of our work in [A.P.~Gantapara~\emph{et~al.}, Phys. Rev. Lett. \textbf{111}, 015501 (2013)], we investigate here the thermodynamic phase behavior of a family of truncated hard cubes, for which the shape evolves smoothly from a cube \emph{via} a cuboctahedron to an octahedron. We used Monte Carlo simulations and free-energy calculations to establish the full phase diagram. This phase diagram exhibits a remarkable richness in crystal and mesophase structures, depending sensitively on the precise particle shape. In addition, we examined in detail the nature of the plastic crystal (rotator) phases that appear for intermediate densities and levels of truncation. Our results allow us to probe the relation between phase behavior and building-block shape and to further the understanding of rotator phases. Furthermore, the phase diagram presented here should prove instrumental for guiding future experimental studies on similarly-shaped nanoparticles and the creation of new materials. 
\end{abstract}

\date{\today}
\maketitle

\section{Introduction}

Material design based on nanoparticle assemblies have been at the focus of materials science over the past decade. In particular, the self-assembly of polyhedral colloidal nanoparticles into functional materials with targeted properties has attracted huge interest. Recent advances in experimental techniques led to the synthesis of a wide variety of polyhedron-shaped particles, such as cubes,~\cite{Sun,Ma2010,Zeng2012,Wang2013} truncated cubes,~\cite{Sun,henzie_self-assembly_2011,Evers2012,Xia2012} truncated octahedra,~\cite{Barnard,henzie_self-assembly_2011} octahedra,~\cite{henzie_self-assembly_2011} tetrahedra,~\cite{Matijevic_0} superballs,~\cite{Rossi} and rhombic dodecahedra.~\cite{Wu,rao2014} In addition to controlled synthesis, the ability to perform self-assembly experiments with these polyhedral particles~\cite{henzie_self-assembly_2011,Bai2012,Zhang2012,Eguchi2012,Quan2012,Choi2012,Zhang2011,Rossi} has made significant strides forward.

This motivated many physicists, mathematicians, and computer scientists to investigate and try to classify the close-packed structures exhibited by these particles. Initially, the focus lay on the prediction of the maximum crystalline packing for faceted particles, as these structures are likely to form upon deposition and evaporation, and also have interesting geometric properties.~\cite{Betke,Murray2000,Torquato_1} Recent extensive investigations of many particle shapes demonstrated the importance of shape for the high-density (close-packed) structures. In particular, De Graaf~\emph{et al.}~\cite{Graaf_PRL} investigated the closed packed structures of 142 convex polyhedra, as well as 17 nonconvex faceted shapes. More recently, Chen~\emph{et al.}~\cite{chen2014} considered over 55000 convex shapes, using theoretical, numerical, and computational methods. 

Advances in computer power and performance have made it possible to perform simulations of these systems with large numbers of particles and opened up the way for a thorough examination of the phase behavior of faceted colloids at finite pressures, \emph{i.e.}, at densities below close packing. Some of the first examples of the importance of studying larger (non-crystalline) assemblies appeared, when it was discovered that there exists a tetrahedron packing into a quasicrystal arrangement with close-packed density much higher than that of spheres.~\cite{Conway2006,Chaikin2010,Torquato_1,Akbari_Glotzer} Simultaneously, the phase diagrams for hard superballs,~\cite{Torquato_8,Ni2012} a family of truncated tetrahedra,~\cite{Glotzer_7} and a family of truncated cubes~\cite{Gantapara2013} were established. The importance of mesophase structures was further underpinned by investigations of space filling polyhedra~\cite{Agarwal}, truncated cubes~\cite{Thapar2014_2} and bifrustums~\cite{Gantapara2014} at an interface, and a large number of polyhedral particles.~\cite{Damasceno2012}

To date, only one experimental investigation of a family of truncated particles has been undertaken. Using a polyol synthesis technique, Henzie~\emph{et~al.}~\cite{henzie_self-assembly_2011} reported the shape-controlled synthesis of monodisperse silver (Ag) nanocrystals including cubes, truncated cubes, cuboctahedra, truncated octahedra, and octahedra. They used these polyhedral particles to study the close-packed crystal structures via sedimentation experiments and simulations. Henzie~\emph{et~al.} created exotic superlattices with potential applications in nanophotonics, photocatalysis, and plasmonics. Their results tested several conjectures on the densest packings of hard polyhedra.~\cite{Betke,Murray2000,Minkowski,Torquato_1} In addition to the close-packed structure studies in the bulk, they also investigated the influence of walls on the sedimented structures.

However, Henzie~\emph{et~al.} did not examine the finite-pressure behavior of the system. At finite pressures the structures that form by self-assembly, may differ substantially from the packings achieved at high (sedimentation and solvent-evaporation) pressures. For instance, simulations of superballs and of truncated cubes exhibited plastic-crystal phases,~\cite{Ni2012,Agarwal,Gantapara2013} while cubes, cuboids and truncated cubes exhibit vacancy-rich simple cubic~\cite{Smallenburg,Marechal2012,Gantapara2013}, and tetrahedra exhibit quasicrystalline mesophases.~\cite{Glotzer_7} In fact, for almost all truncated particle shapes studied thus far, the finite-pressure phases and those formed under close-packed conditions differ substantially.~\cite{Damasceno2012}

In this manuscript we investigate the finite-pressure behavior of particles similar to those considered by Henzie~\emph{et~al.} Here, we present a thorough investigation of the phase behavior of a family of truncated hard cubes, which interpolates smoothly between cubes and octahedra (the mathematical dual of the cube) \emph{via} cuboctahedra. We describe in detail the different phases, as well as the nature of the phase transitions between these phases. This work is an extension of our previous investigation of these systems, see Ref.~\rcite{Gantapara2013}, and makes several minor updates on our previous results. In the present paper, we put additional emphasis on the analysis of the plastic crystal or rotator phases and the computational details. 

We used Monte Carlo simulation studies and free-energy calculations to establish the phase diagram for this system. This diagram exhibits a remarkably rich diversity in crystal structures that show a sensitive dependence on the particle shape. Changes in phase behavior and crystal structures occur even for small variations in the level of truncation. This is an unexpected result, since the particle shape varies smoothly from that of a cube to that of an octahedron by truncation. We also observed that for specific levels of truncation the particles possess an equation of state (EOS) that exhibits three distinct crystal phases as well as an isotropic fluid phase. 

In addition, we found that for close-to-cubic particles that form a vacancy-rich simple cubic phase the equilibrium concentration of vacancies increases at a fixed packing fraction $\phi$ upon increasing the level of truncation, see Gantapara~\emph{et al.}~\cite{Gantapara2013} The vacancy concentrations for truncated cubes for small truncations are in agreement with the vacancy concentration of perfect cubes.~\cite{Smallenburg} However, our results differ from those obtained by Monte Carlo simulations for parallel cuboids, where the vacancy concentration remains constant, when the shape is varied from a perfect cube to a sphere \emph{via} rounded cubes (so-called cuboids).~\cite{Marechal2012} 

Furthermore, we analyzed the orientation distribution properties of particles in the plastic-crystal phases observed in the truncated cubes phase diagram. The orientation distribution function of plastic crystals of hard anisotropic particles is shown to be highly anisotropic and strongly peaked for specific orientations. Based on our results, we present a grouping of particles with different asphericity $A$ values according to their cubatic order $S_4$ near (plastic-)crystal-fluid transition regions. We find that particles with asphericity $A<0.1$ exhibit plastic-crystal phases with cubatic order $S_4$ as low as $S_4\approx0.1$, comparable to the cubatic order values in the isotropic fluids. Our results show that the cubatic order in a system is inversely proportional to the number of preferential orientations of the truncated cubes in the bulk plastic crystal phase.
 
The remainder of the paper is organized as follows. We first present our simulation model in Section~\ref{sec:sm}. We discuss the simulation methods as well as the order parameters and correlation functions used in our analysis of the phase behavior in Section~\ref{sec:smethods}. The results are presented in Section~\ref{sec:results}. In particular, the close-packed structures are presented in Section~\ref{subsub:FBMC} and ~\ref{sub:family}, followed by a discussion of the full phase diagram in Section~\ref{sec:pd}. In Section~\ref{sub:plastic_crystals}, we analyze the orientation distributions of particles in the various plastic-crystal phases observed in the phase diagram. Finally, we discuss the results and draw conclusions in Section~\ref{sec:conclusions}.

\section{\label{sec:sm}Simulation Model}

The particles that we investigated are completely specified by the level of truncation of a perfect cube, which we denote by $s \in [0,1]$, and the volume of the particle. We define our family of truncated cubes using a simple mathematical expression for the location of the vertices. The line segments that connect these vertices can only be assigned in one (unique) way to obtain a truncated cube. The vertices of a truncated cube may be written as a function of the shape parameter $s \in [0,1]$:
\begin{equation}
\label{INeq:trunc} \{ \mathbf{v}(s) \} = \left\{ \begin{array}{l} \left( 1 - \frac{4}{3}s^{3} \right)^{(-1/3)} \mathcal{P}_{D} \left( \pm \left(\frac{1}{2} - s \right), \pm \frac{1}{2}, \pm \frac{1}{2} \right)^{T} \\ \\ \quad s \in \left[ 0,\frac{1}{2} \right] \\ \\
 \left( \frac{4}{3} - 4\lambda^{3} \right)^{(-1/3)} \mathcal{P}_{D} \left( \pm (1 - \lambda), \pm \lambda, 0 \right)^{T} \\ \\ \quad \lambda \equiv 1 - s \in \left[ 0,\frac{1}{2} \right] \end{array} \right. ,
\end{equation}
where $\mathcal{P}_{D}$ is a permutation operation that generates all permutations of each element in the sets of 8 and 4 vertices spanned by the $\pm$-operations, respectively. The duplicate vertices that are a consequence of this definition are removed after letting $\mathcal{P}_{D}$ act. The `$T$' indicates transposition. The prefactors ensure that the truncated cubes are normalized to unit volume. Several Platonic and Archimedean solids are members of this family: $s = 0$ a cube, $s = (2 - \sqrt{2})/2 \approx 0.292893$ a truncated cube, $s = 1/2$ a cuboctahedron, $s = 2/3$ a truncated octahedron, and $s = 1$ an octahedron; these are depicted in Fig.~\ref{fig:dense}a. 

\section{\label{sec:smethods}Simulation Methods}

\subsection{\label{sec:ord}Order Parameters and Correlations Functions}

In this subsection we describe different order parameters that we used to quantify the positional and orientation order of particles in our isothermal-isobaric Monte Carlo simulations (also called \emph{NPT} simulations; fixed pressure $P$, temperature $T$, and number of particles $N$) of the truncated cubes. These order parameters play a crucial role in identifying different phases exhibited by the truncated cubes. Truncated cubes have cubatic symmetry. To quantify the orientation order for these particles the cubatic order parameter $S_4$ is appropriate as was shown in earlier simulation studies on cubatic particles.~\cite{batten2010,Ni2012} The cubatic order parameter is defined as 
\begin{equation}
\label{eq:S4} S_4 = \max_{\mathbf{n}}{\left\{\frac{1}{14N}\sum_{i,j}{\left(35|\mathbf{u}_{ij} \cdot \mathbf{n} |^4 - 30|\mathbf{u}_{ij} \cdot \mathbf{n} |^2 +3 \right)}\right\}},
\end{equation}
where $N$ is the number of particles as above, $\mathbf{u}_{ij}$ is the unit vector along the main axis ($j\in\{ x,y,z\}$) of particle $i$ and {\bf n} is the unit vector for which $S_4$ is maximized. $S_4$ values range from $0$ for a completely disordered system to $1$ for perfect crystals.

To investigate the structural correlations in the particle orientations we use an orientation correlation function $g_4(r)$ defined as
\begin{equation}
\label{eq:g4} g_4(r)= \frac{3}{14}\left< 35 [\mathbf{u}_{aj}(0).\mathbf{u}_{bj}(r)]^4 - 30 [ \mathbf{u}_{aj}(0).\mathbf{u}_{bj}(r)]^2 +3 \right> , 
\end{equation}
where $\left<.\right>$ denotes the ensemble average over all the particle axes $j\in\{x,y,z\}$ and particle pairs $a$ and $b$. For more details about the definitions and computation of these order parameters we refer the reader to Batten~\emph{et al.}~\cite{batten2010}

To determine the translational order in the system we use the radial distribution function $g(r)$ defined as 
\begin{equation}
\label{eq:g2} g(r)=\frac{1}{\rho^2}\left< \sum_{i=1}^N \sum_{j\not=i }^N \delta \left( {\bf r}-{\bf r}_{i} \right)\delta \left( {\bf r^\prime}-{\bf r}_{j} \right)\right>,
\end{equation} 
with $r=\left|\mathbf{r}-\mathbf{r}^\prime\right|$, $\delta (x)$ is the usual Kronecker $\delta$-function, ${\bf r}_{i}$ and ${\bf r}_{j}$ are the positions of the $i^{\text{th}}$ and $j^{\text{th}}$ particle, respectively, and $\rho =N/V$ is the number density of the system. The radial distribution function together with the order parameters are useful to distinguish plastic crystal from crystal and isotropic fluid phases.

\subsection{\label{ch2:sub:Free}Free-Energy Calculations and Confining Potentials}

We obtained the dimensionless free energy per particle $f=\beta F/N$ as a function of packing fraction $\phi=N v_p /V$, with $v_p$ the particle volume ($\phi=\rho$, since $v_p$ is the unit of volume in this manuscript), for the entire density range by thermodynamic integration~\cite{Frenkelbook:UMS} over the equation of state (EOS), from reference density $\rho_0$ to the density of interest $\rho$:
\begin{equation}
\label{eq:thermint} f(\rho) = f(\rho_0) + \int_{\rho_0}^{\rho}{\frac{\beta P(\rho')}{\rho'^2}\, \mathrm{d}\, \rho'}.
\end{equation}
Here, $f(\rho_0) \equiv \beta \mu(\rho_0) - \beta P(\rho_0)/\rho_0$ is the reduced Helmholtz free energy per particle at density $\rho_0$, with $\beta=1/k_\mathrm{B} T$, with $T$ the temperature and $k_\mathrm{B}$ the Boltzmann constant, $\mu(\rho_0)$ the chemical potential, and $P(\rho_{0})$ the pressure. The Helmholtz free energy at reference density $\rho_0$ was obtained as follows.

\begin{enumerate}
\item In the fluid phase we used Widom's particle insertion method~\cite{Widom1963} to obtain the free energy. This method was employed at relatively low densities to obtain small error bars. We performed the calculations at $\phi\approx 0.2$. We note that there were no finite-size effects within the computational accuracy for the particle insertion method.

\item In the crystal phase we used the Einstein integration method.~\cite{Frenkelbook:UMS,Frenkel1984,Noya2007} The reduced Helmholtz free energy per particle $f=\beta F/N$ of a crystal is given by:
\begin{eqnarray}
\nonumber f(\rho) & = & f_{\mathrm{Einst}}(\lambda_{\max}) - \\
\label{eq:freecrys} & & \frac{1}{N}\int_{0}^{\lambda_{\max}}{\mathrm{d}\lambda\left\langle\frac{\partial \beta U_{\mathrm{Einst}}(\lambda)}{\partial\lambda}\right\rangle},
\end{eqnarray}
where $f_{\mathrm{Einst}}$ denotes the reduced free energy per particle of the ideal Einstein crystal, which is given by:
\begin{eqnarray}
\nonumber \qquad f_{\mathrm{Einst}}(\lambda_{\max}) & = & - \frac{3(N-1)}{2N}\log{\left(\frac{\pi}{\lambda_{\max}}\right)} + \\
\nonumber & & \log{\left(\frac{\Lambda_{t}^{3}\Lambda_r}{v_p} \right)}+ \frac{1}{N}\log{\left(\frac{v_p}{VN^{1/2}} \right)} -\\
\nonumber & & \frac{1}{N}\log \left\{ \frac{1}{8\pi^2} \int\mathrm{d}\theta \sin{(\theta)}\mathrm{d}\phi\mathrm{d}\chi \right. \times \\
\nonumber & & \left. \exp{\left[ - \frac{\lambda_{\max}}{k_{B}T}(\sin^2{\psi_{ia}} + \sin^2{\psi_{ib}})\right]} \right\}. \\
\label{eq:freein} & &
\end{eqnarray}
$U_{\mathrm{Einst}}(\lambda)$ denotes the harmonic potential that fixes the particles to the respective Einstein lattice positions and orientations:
\begin{eqnarray}
\nonumber \beta U_{\mathrm{Einst}}(\lambda) &=& \lambda \displaystyle \sum_{i=1}^{N} [(\mathbf{r}_i-\mathbf{r}_{i,0})^2/v_p^{2/3} \\
\label{eq:einst_pot} &+& (\sin^2\psi_{ia} + \sin^2\psi_{ib})] ,
\end{eqnarray}
with $(\mathbf{r}_i-\mathbf{r}_{i,0})$ the displacement of particle $i$ from its position in the ideal Einstein crystal. The angles $\psi_{ia}$ and $\psi_{ib}$ are the minimum angles between vectors, $\mathbf{a}$ and $\mathbf{b}$, describing the orientations of the particles in the ideal Einstein crystal and the equivalent vectors that describe the orientation of the particle in the actual crystal, respectively. The translational and rotational thermal wavelengths $\Lambda_t$ and $\Lambda_r$ in Eq.~\cref{eq:einst_pot} were set to unity in our calculations. When $\lambda$ is large the translational and orientation displacements of the particles are frozen, while at lower $\lambda$'s the particles freely displace and rotate, exploring the underlying degeneracy coming from the symmetry of the particle itself. 

We typically used system sizes of 700 to 1,500 particles to compute the free energies for the (plastic) crystal phases. We found that finite-size scaling (FSS) was only necessary in the octahedron regime, \emph{i.e.}, $s\approx1$, to establish the phase diagram. For such high levels of truncation the free-energy differences between the various phases at coexistence proved to be very small, see Ni~\emph{et al.}~\cite{Ni2012} For the other phase transitions the free energies obtained without FSS proved to be sufficient to accurately determine the phase boundaries.

\item For the free-energy calculations of a plastic-crystal (rotator) phase, we followed the approach of Fortini~\emph{et al.},~\cite{Fortini2005} who introduced a method, which allows for a continuous transition from a non-interacting plastic-crystal to an interacting plastic-crystal phase of hard truncated cubes. We used a tunable soft-to-hard interaction potential between the particles
\begin{equation}
\qquad \quad \varphi(i,j) = \left\{\begin{array}{cl}
\gamma[1-A(1+\zeta(i,j))] & \mathrm{if}\;\zeta(i,j) < 0 \\
0 & \mathrm{otherwise}
\end{array} \right. .
\end{equation}
Here, $\zeta(i,j)$ is the overlap potential defined in Donev~\emph{et al.,}~\cite{Donev2012} which is negative when two particles $i$ and $j$ overlap and positive otherwise. The integration parameter $\gamma$ runs from $0$ (noninteracting) to $\gamma_{\max}$, for which the system interacts fully. In our calculations we set $A=0.9$ following Marechal~\emph{et al.}~\cite{Marechal2010} The dimensionless Helmholtz free energy per particle in the plastic crystal is given by:
\begin{eqnarray}
\nonumber \qquad f(\rho) & = & f_{\mathrm{Einst}}(\lambda_{\max}) - \\
\nonumber & & \frac{1}{N}\int_{0}^{\lambda_{\max}}{\mathrm{d}\lambda\left\langle\frac{\partial \beta U_{\mathrm{Einst}}(\lambda)}{\partial\lambda}\right\rangle_{\gamma_{\max}}} + \\
\nonumber & & \frac{1}{N}\int_{0}^{\gamma_{\max}} \mathrm{d}\gamma \left\langle \frac{\partial \sum_{i\neq j}^{N}{\beta \varphi(i,j)}}{\partial\gamma} \right\rangle_{\lambda_{\max}}. \\
\label{eq:free_rot} & &
\end{eqnarray}
\end{enumerate}

\section{\label{sec:results}Results}

\subsection{\label{subsub:FBMC}Determining the Close-Packed Structures}

\begin{figure*}[!hbt]
\begin{center}
\includegraphics[scale=1]{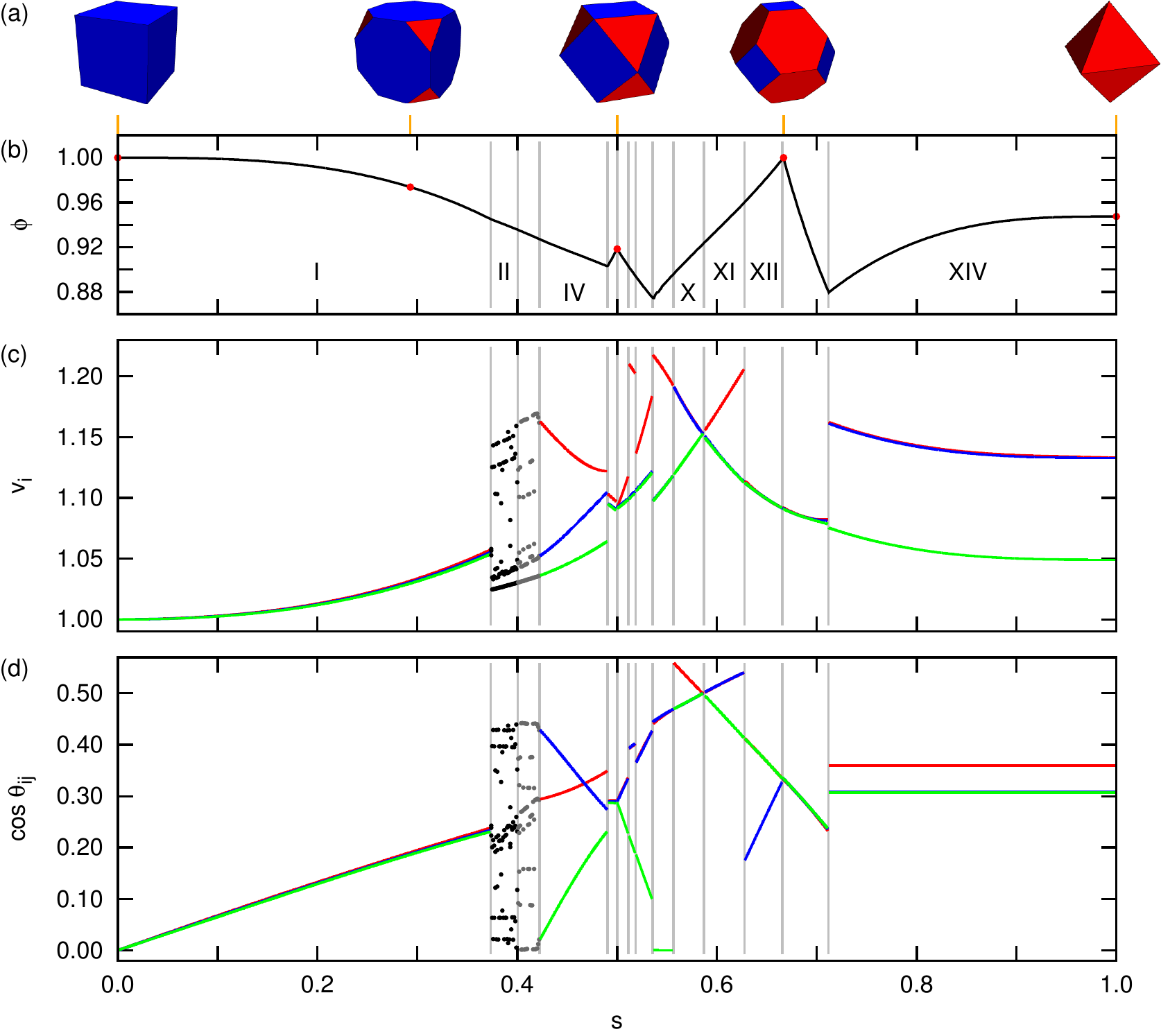}
\end{center}
\caption{\label{fig:dense} (a) Five examples of truncated cubes (Platonic and Archimedean solids only) for levels of truncation $s$ corresponding to the orange lines: $s = 0$ a cube, $s = (2 - \sqrt{2})/2 \approx 0.293$ a truncated cube, $s = 1/2$ a cuboctahedron, $s = 2/3$ a truncated octahedron, and $s = 1$ an octahedron. (b) The packing fraction $\phi$ for the close-packed structures as a function of $s$. The values for the five solids shown in (a) are given by red dots. (c) The length $v_{i}$ ($i = 1$, $2$, and $3$) of the three lattice vectors, indicated in red, green, and blue, that span the unit cell of the densest crystal structure as a function of $s$. Not every line is clearly visible, since there is some overlap. In the region where the black and gray dots are used ($s \in (0.37,0.40]$ and $s \in  (0.40,0.42]$), there appears to be a degeneracy in the crystal structures, as explained in the text. (d) The cosine of the angles $\theta_{ij}$ ($i < j = 1$, $2$, and $3$) between the three vectors that span the unit cell as a function of $s$. Gray vertical lines partition the $s$-domain into 14 pieces with a `different' crystal structure, based on the discontinuities shown in the $v_{i}$ and $\cos \theta_{ij}$ results. These regions are numbered with roman numerals in (b); only those regions large enough to accommodate a label are numbered, but the numbering can be continued from left to right in the unnumbered regions.}
\end{figure*}

The simulations by which the close-packed structures were derived, are based on the floppy-box Monte Carlo (FBMC) method~\cite{lfilion2009,Graaf2012} in combination with the separating-axis-based overlap algorithm.~\cite{MSA} We obtained the densest crystal structure and the corresponding packing fraction $\phi$ as a function of the level of particle truncation $s$ by considering 1,000 equidistant points in $s \in [0,1]$. For each point we prepared systems of truncated particles in a dilute phase, typically with packing fraction $\phi \approx 0.001$. We increased the reduced pressure in $100$ steps according to a geometric series from $p = 1$ to $p \approx 10^{5}$ over $4 \times 10^{6}$ Monte Carlo (MC) cycles in order to compress these systems to a high-density crystalline state. This pressure increase was typically applied a total of 1,000 times for $N = 1$ particles in the unit cell and for each shape. We restricted ourselves to $N = 1$ particles in the unit cell, because the truncated cubes are all centrosymmetric. We only considered $N = 2$, $\dots$, $6$ for $14$ conveniently chosen values of $s$, located in the center of the regions indicated in Fig.~\ref{fig:dense}, as will be justified shortly. For these $N > 1$ systems we obtained roughly the same value of $\phi$ and also the same crystal structures. The densest crystal-structure candidate was selected and allowed to compress further for another $10^{6}$ MC cycles at $p = 10^{6}$ to achieve 5 decimals of precision in $\phi$. In practice, these final cycles of compression did not improve the packing fraction substantially. Figure~\ref{fig:dense}b shows $\phi$ as a function of $s$. Note that the packing fraction `curve' is continuous, but has discontinuities in its first derivative. To double check our result, we considered another set of FBMC runs. For these we took several of the 1,000 densely-packed crystals as our initial configuration and varied $s$ around the selected points at high pressure to study the evolution of the crystal structure. Steps of $10^{-5}$ in $s$ were employed and for each step the system is expanded to remove any overlaps, before re-compressing it at $p \approx 10^{5}$. The packing fractions we obtained showed good correspondence with our original result, but this correspondence failed for a transition between two crystal structures. The consecutive method would often become stuck in the lower density structure that corresponded to the morphology of the crystal phase it came from.

The unit cell for $N = 1$ truncated cubes can be specified by three vectors $\mathbf{v}_{i}$ ($i = 1$, $2$, $3$) that are implicitly $s$ dependent. The structure spanned by these three vectors can also be described by the length $v_{i} = \vert \mathbf{v}_{i} \vert$ of the vectors and the angles $\theta_{ij}$ ($i < j = 1$, $2$, $3$) between them. Note that we ignored the orientation of the particle with respect to the unit cell here. In order to give an unbiased comparison of the different vectors we used lattice reduction~\cite{Likos} to ensure that for each unit cell the surface-to-volume ratio is minimal. These results are shown in Fig.~\ref{fig:dense}(c,d), respectively. By analyzing the $v_{i}$ and $\theta_{ij}$, as well as the location of the kinks in the $\phi$-curve, we were able to partition the $s \in [0,1]$ domain into 14 distinct regions. This is the reason behind our choice of 14 verification points for $N > 1$ simulations. Below we discuss the crystal structures in the different regions and the way these regions can be grouped.

\subsection{\label{sub:family}Properties of the Close-Packed Structures}

\begin{figure*}[!hbt]
\begin{center}
\includegraphics[width=\textwidth]{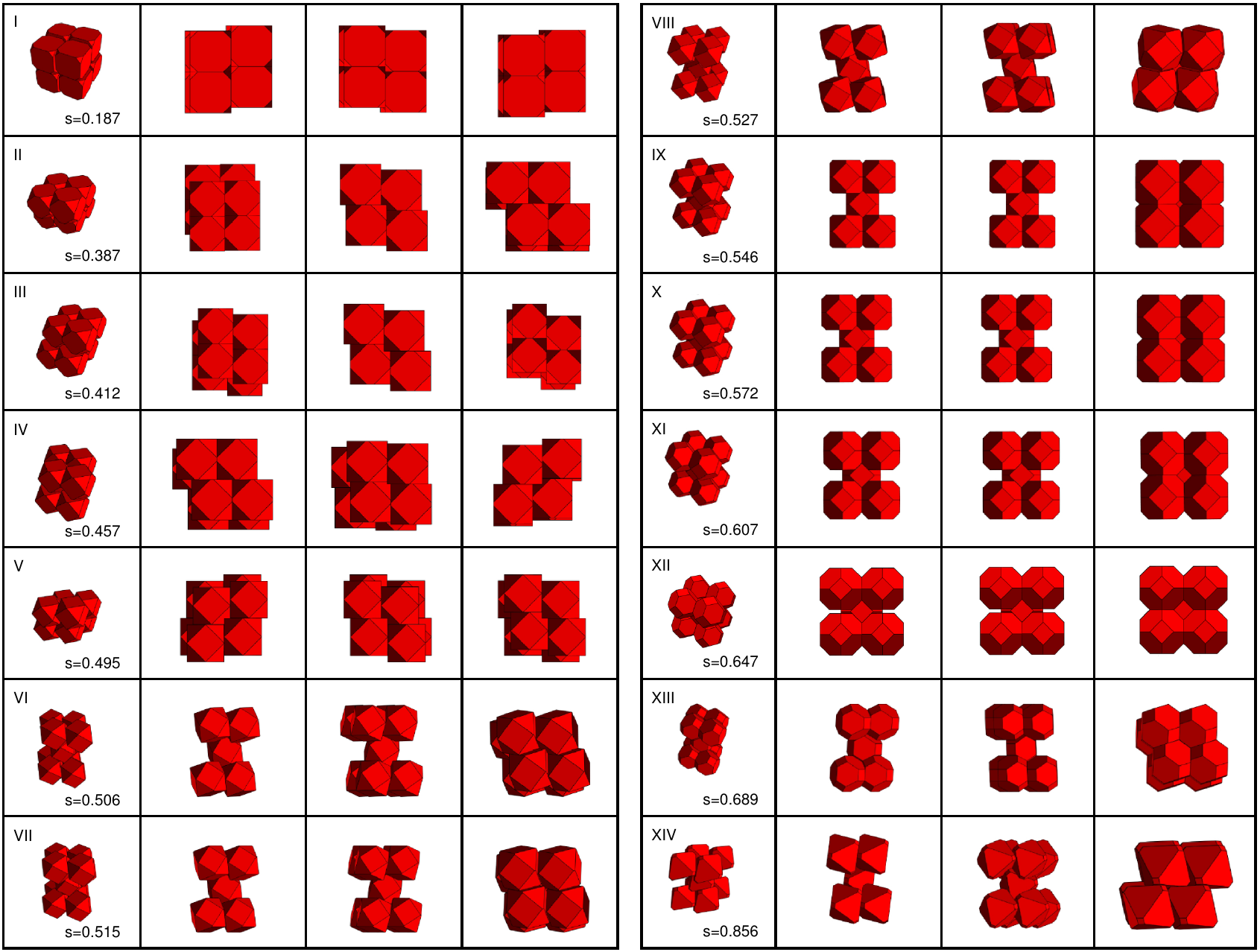}
\end{center}
\caption{\label{fig:structs} Visual representations of the crystal structures obtained for the first 7 regions (left) and the last 7 regions (right) of Fig.~\ref{fig:dense}. From left to right each entry (row) contains a bird's eye view, the front view, the side view, and the top view of this structure. The Roman numeral in the top-left corner gives the relevant domain in Fig.~\ref{fig:dense}. The truncation parameter $s$ for these structures is given in the bottom-right corner of the first panel.}
\end{figure*}

Figure~\ref{fig:structs} shows the crystal structure in the center of each of the 14 regions that we found in Fig.~\ref{fig:dense}. There is a strong difference between the domains $s < 1/2$ and $s > 1/2$. Geometrically the cuboctahedron $(s = 1/2)$ is the transition point between shapes which have a more cube-like nature and shapes which have a more octahedron-like nature. It is therefore not surprising that the crystal structures in the two regions ($s < 1/2$ and $s > 1/2$) appear to have a deformed simple cubic symmetry and a deformed body-centered tetragonal symmetry, respectively. We illustrate this in Fig.~\ref{fig:structs}, where we show the most orthorhombic unit cell: $N = 1$ for $s < 1/2$ and $N = 2$ for $s > 1/2$. A remarkable result is the stability of the Minkowski crystal,~\cite{Minkowski} which is the densest-packed Bravais-lattice structure for octahedra,~\cite{Betke} under variations in $s$. For all $s \in [0.71,1]$, we find a Minkowski structure in the dense-packed limit, which can be inferred from the horizontal $\cos \theta_{ij}$ lines in Fig.~\ref{fig:dense}d. The scaled length of the vectors $v_{i}\phi^{-1/3}$ is also constant on this domain.

Let us now examine the crystal structures in the 14 regions identified by the discontinuities in the vectors of the unit cell. In literature it has become commonplace to assign atomic equivalents to structures observed in simulations or experiments. For example, this is done for binary mixtures of spheres,~\cite{lfilion2009,LFilion} a family of truncated tetrahedra,~\cite{Glotzer_7} several faceted particles,~\cite{Damasceno2012,chen2014} and systems of nanoparticles.~\cite{evers_observation_2009,evers_entropy-driven_2010} We attempted to follow suit by determining the symmetry group of the structures in Fig.~\ref{fig:structs} using \emph{FindSym}~\cite{FindSym} and by subsequently assigning an atomic equivalent.~\cite{CrysStruc} However, we found that a description in terms of atomic equivalents inadequately captures the richness in crystal structure, since particle orientation is not taken into account. Moreover, for many of our structures we are unable to determine a nontrivial space group using \emph{FindSym}. We therefore resorted to visual analysis and we used this to group the $14$ regions in Fig.~\ref{fig:dense} based on similarities between the respective structures. 

In the supplement to Gantapara~\emph{et~al.}~\cite{Gantapara2013} this grouping was originally discussed and it was subsequently commented upon in the work of Chen~\emph{et~al.}~\cite{chen2014} In the latter, the fact that we identified 14 distinct regions in the packing fraction was mistakenly interpreted to mean that these regions all had different crystal structures. Here, we discuss the comparison and show how our original grouping of regions for the different crystal structures corresponds and differs from the one provided by Chen~\emph{et~al.}

\begin{figure}[!h]
\begin{center}
\includegraphics[scale=1.0]{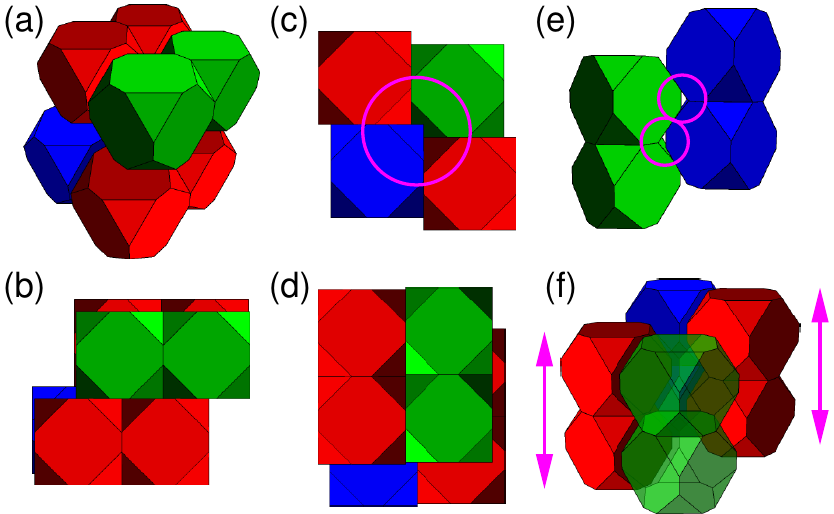}
\end{center}
\caption{\label{fig:sliding} A visual representation of the degenerate crystal structure; we consider the value $s = 0.387$ in this figure. Pairs of truncated cubes, for which the octahedral faces are aligned (columns), are color-coded. Different viewpoints are displayed for a piece of crystal consisting of 8 particles: (a) Bird's eye view, (b) side view, (c), front view, and (d) top view of this structure. In (c) we use a magenta circle to indicate that the blue column is interlocking with the green column in a diagonal way. (e) A diagonal view of the crystal structure, where the red columns have been removed. Magenta circles show the interlocking. (f) The two red columns are \emph{not} interlocking with the blue and green column, allowing for freedom of motion in the direction of the magenta arrows. The green column is made translucent to better illustrate the properties of this crystal structure.}
\end{figure}

\begin{enumerate}
\item \textbf{I} In this region ($s \in [0.00,0.37]$) we obtained a continuous and uniform distortion of the simple cubic structure for cubes. For $s = 0$ the particles form a simple cubic (SC) crystal, which has the same morphology as $\alpha$Po ($\alpha$-Polonium).~\cite{CrysStruc} The uniformly distorted simple cubic (UDSC) structure we found for $s > 0$ is similar to that of $\beta$Po.~\cite{CrysStruc} We verified this distorted quality for values as low as $s = 10^{-5}$. This region corresponds to $\rho_{7}$ in Chen~\emph{et~al.}

\item \textbf{II~\&~III} For these two regions ($s \in (0.37,0.40]$ and $s \in (0.40,0.42]$) we found that there is a degeneracy in the crystal-structure candidates that achieve the densest-known packing. Although certain structures appear favored over others, there is no clear relation between the structure and $s$. However, the packing fraction $\phi$ of the close-packed crystals is continuous in these regions. 

The observed degeneracy can be explained by the formation of sheets consisting of diagonally-interlocked columns, which can slide up or down (in the direction of the columns) with respect to each other, as shown in Fig.~\ref{fig:sliding} for $s = 0.387$. The truncated cubes are arranged in a distorted simple cubic (DSC) crystal lattice, where the particles form columns that are interlocked in a diagonal way. These structures are referred to as mono-interlocking distorted simple cubic (MI-DSC) crystals. This diagonal interlocking together with the close-packing condition, prevents lateral motion in the plane normal to the column's direction. However, since the system is not fully interlocked, motion in the direction of the columns is possible for the diagonally interlocked sheets. 

The observed degeneracy is different from the degeneracy that occurs in structures consisting of cubes or hexagonal prisms for instance, since such systems allow lateral freedom of movement of columns or (perpendicular to the columns) of sheets of aligned particles. That is, there is possible freedom of motion in three directions, albeit not necessarily at the same time. The interlocking nature of the MI-DSC phase only allows for movement in one direction only, namely parallel to the columns, which may lead to strong rheological differences between this structure and, \emph{e.g.}, the SC structure for cubes. This grouping corresponds to region $\rho_{2}$ in Chen~\emph{et~al.}

\item \textbf{IV} For this region ($s \in (0.42,0.49]$) we find a DSC phase that is interlocking in two directions: a bi-interlocking DSC (BI-DSC) phase. For each instance of interlocking two degrees of translational motion are frozen out. This implies that the BI-DSC structure is completely fixed, which is confirmed by the unicity of the $v_{i}$ and $\theta_{ij}$ results in Fig.~\ref{fig:dense}(c,d). This region corresponds to $\rho_{5}$ in Chen~\emph{et~al.}

\item \textbf{V} In this region ($s \in (0.49,0.50]$) we observed a tri-interlocking DSC (TI-DSC) phase. This region corresponds to $\rho_{0}$ in Chen~\emph{et~al.}

\item \textbf{VI~-~VIII} Here ($s \in (0.50,0.51]$, $(0.51,0.52]$, and $(0.52,0.54]$) we found structures that are best described by a distorted body-centered tetragonal (DBCT) structure. The truncated cubes in these crystals are not aligned with the axes of the unit cell. It is unclear to what extent structures in regions VI, VII, and VIII are the same. The smooth flow of the $\phi$-curve (Fig.~\ref{fig:dense}b), as well as their appearance, s implies continuity, but the jumps in the values of $v_{i}$ and $\theta_{ij}$ [Fig.~\ref{fig:dense}(c,d)] suggest otherwise. This grouping corresponds to region $\rho_{4}$ in Chen~\emph{et~al.} in which the subregions are considered to be the same.

\item \textbf{IX~-~XII} These structures ($s \in (0.54,0.56]$, $(0.56,0.59]$, $(0.59,0.63]$, and $(0.63,0.67]$) have a body-centered tetragonal (BCT) morphology, for which the particles are aligned with the lattice vectors of the unit cell. Originally, we had assigned region \textbf{XII} to a separate structure. Chen~\emph{et~al.} correctly pointed out that regions IX~-~XII belong to the same crystal structure, namely $\rho_{6}$ in their notation. It should be further pointed out that in this region the BCT structure smoothly deforms into a BCC structure for $s = 2/3$, by increasing $s$. 

\item \textbf{XIII} This DBCT structure ($s \in (0.67,0.71]$) is different from the DBCT structures in regions VI~-~VIII, since the particles appear to be aligned with the lattice vectors of the unit cell. Moreover, crystals in this region are unusual, since there are large `voids' in the structure. That is, for all other structures we found that the largest facets of a particle are always in contact with a similar facet of another particle. This is not the case here, because there is a substantial gap between some of the hexagonal facets. Chen~\emph{et~al.} assign our region XIII to their $\rho_{1}$

\item \textbf{XIV} The Minkowski crystal of region XIV ($s \in (0.71,1.00]$) is also noteworthy. It is the only structure which does not undergo any reorganization upon varying the level of truncation. It is worthwhile to study the origin of this apparent stability, which sharply contrasts with the immediate distortion found around $s = 0$. However, this goes beyond the scope of the current investigation. This region corresponds to region $\rho_{3}$ in Chen~\emph{et~al.}
\end{enumerate}

In conclusion, our visual-inspection-based grouping of the 14 regions of Fig.~\ref{fig:dense} leads to 8 distinct crystal structures being identified. This grouping is the same as the one specified in Chen~\emph{et~al.},~\cite{chen2014} after making one correction to our previous finding.~\cite{Gantapara2013} 

\subsection{\label{subsub:eos}Equations of State and Mesophase Structures}

\begin{figure*}[!htb]
\begin{center}
\includegraphics[scale=1]{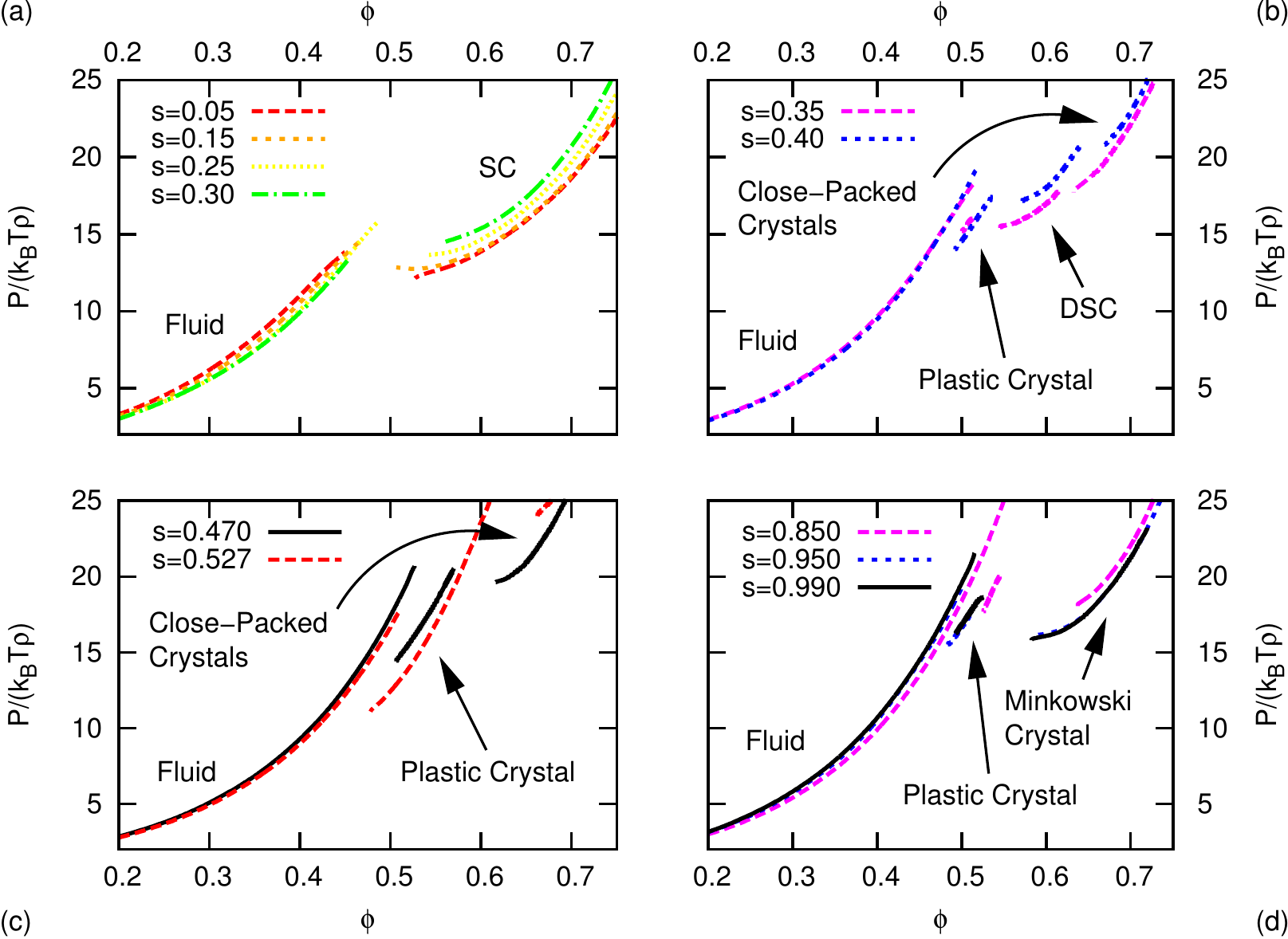}
\end{center}
\caption{\label{fig:eq_compress}The equations of state (EOSs) for a selection of truncation values $s$. The reduced pressures $P/(k_B T \rho)$ is plotted as a function of packing fraction $\phi$, where $P$ is the pressure of the bulk system, $\rho$ is the number density of the system, and $k_B T$ is the thermal energy. The EOSs are grouped into four sets based on their phase behavior.} 
\end{figure*}

We used the close-packed crystal structures obtained from the FBMC calculations as initial configurations for variable-box-shape isothermal-isobaric ($NPT$) Monte Carlo simulations, to study the phase behavior at intermediate pressures. Initial configurations of $300$ to $600$ particles were prepared and melted to determine the equations of state (EOSs) for the various phases. Typical equilibration times were around $1.2 \times 10^{6}$ Monte Carlo sweeps (MCS) and the production times around $2 \times 10^{6}$ MCS. One MCS is defined as $N$ Monte Carlo trial moves (translation, rotation, volume change, or deformation of the box, respectively), where $N$ is the number of particles in the system. We sampled the lattice vectors, as well as the average positions and orientations of the particles as a function of packing fraction and for fixed truncation parameter $s$. The sampling was done on an interval of 100 MCS to avoid correlated configurations. Using these results we set up regular $NPT$ simulations (possibly with a triclinic box shape) to more accurately sample the EOSs for all phases with larger system sizes of 1,000 to 2,000 particles, including the liquid phase. 

In Fig.~\ref{fig:eq_compress} we show the EOS obtained from our FBMC simulations as a function of the packing fraction. We show the EOSs only for selected shapes. The liquid EOS branches were obtained by compressing dilute systems ($\phi\approx0.1$) while the crystalline branches of the EOS were obtained by melting the close-packed structures. We grouped the EOSs on the basis of their phase behavior. EOSs for truncated cubes with truncation $s\leq0.30$ are shown in Fig.~\ref{fig:eq_compress}a. These systems exhibit an isotropic liquid phase and a simple cubic phase separated by a first-order phase transition. During our $NPT$ compression runs we observed that these systems crystallize easily with relatively little hysteresis compared to systems with $s>0.7$. In Fig.~\ref{fig:eq_compress}b we show EOSs for $s=0.35 \text{ and }0.40$. These two shapes, surprisingly, exhibit one isotropic phase and three crystalline phases. The rest of the EOSs in Fig.~\ref{fig:eq_compress}(c,d) show three phases: liquid, plastic crystal and crystalline phase. More details about the phase behavior and individual (plastic-)crystalline phases of these systems will be given in Section~\ref{sec:pd}. These  EOSs were used to calculate the Helmholtz free energies at different packing fractions using thermodynamic integration as explained in Section~\ref{ch2:sub:Free}.

\subsection{\label{sub:averaging} Mesophase Lattice Vectors}

\begin{figure*}[!htb]
\begin{center}
\includegraphics[scale=1]{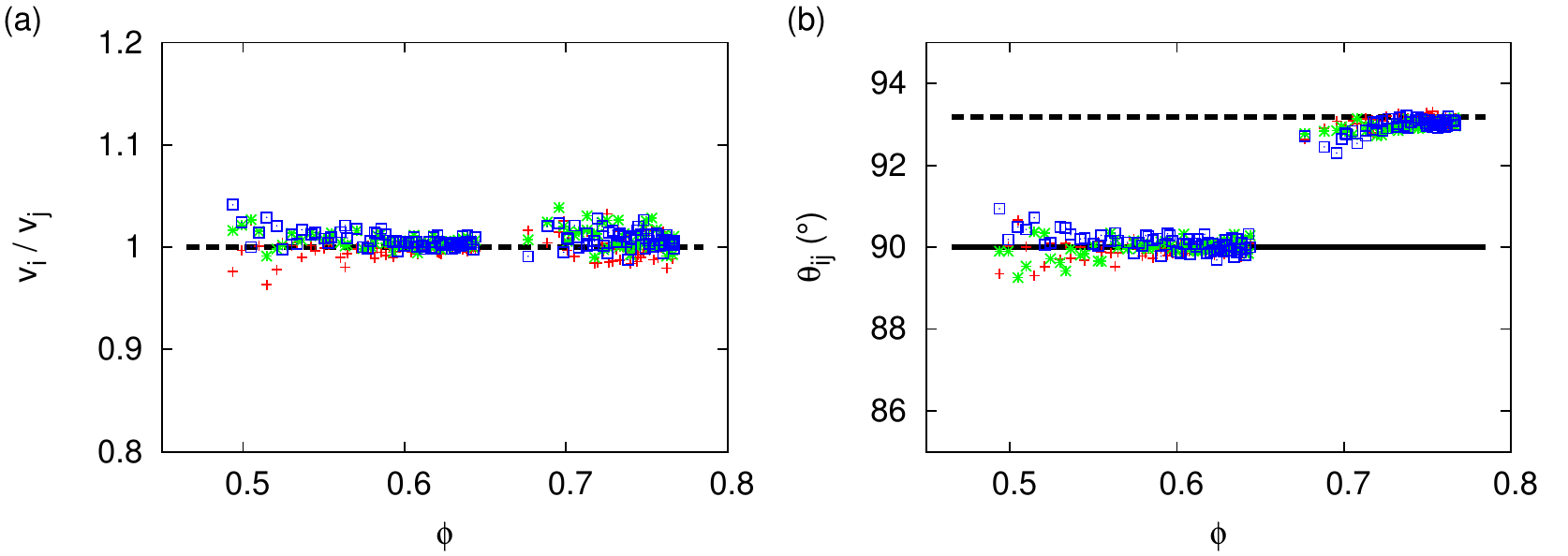}
\end{center}
\caption{\label{fig:lat_vec_s0p750} (a) The ratio of the averaged lattice vectors $v_i$ and $v_j$. (b) The angles $\theta_{ij}$ between these vectors. Both are shown as a function of the averaged packing fraction $\phi$, which was obtained from our \emph{NPT} simulations for $s=0.750$. The indices $i$, $j$ run over all the $x$, $y$, and $z$ components of the box as described in Section~\ref{subsub:FBMC}. The $\theta_{ij}$ values are given in degrees. The black dashed lines correspond to values extracted from the close-packed Minkowski crystal. The black solid line in (b) corresponds to the value of $\theta_{ij}$ for a BCC lattice.} 
\end{figure*}

Before we turn our attention to the phase diagram, we explain how the \emph{NPT} data was used to compute free energies and to determine the crystal structure of the mesophases. To compute these quantities, we determined the inherent ideal lattice at each pressure or packing fraction. This was accomplished by averaging the box vectors and the angles between them during the $NPT$ simulations at each given pressure. Using these averaged quantities we reconstructed an ideal lattice. Visual inspection of the ideal lattice allowed us to determine the crystal structure. We also used the ideal lattice in the free-energy calculations as the reference Einstein crystal. 

To illustrate the averaging procedure, we show the ratio of the lattice lengths $v_{i}/v_{j}$ and the lattice angles $\theta_{ij}$ as a function of packing fraction $\phi$ for $s=0.750$ in Fig.~\ref{fig:lat_vec_s0p750}. The dots in the plots represent the average values from the $NPT$ simulations at each pressure, while the thick lines represent lattice vectors and their angles from the close-packed structure. For $s=0.750$, the close-packed structure is the Minkowski lattice. For the Minkowski lattice, $v_i/v_j=1$ and $\theta_{ij} \approx 93.1847\degree$. A BCC lattice is defined by $v_i/v_j =1$ and $\theta_{ij}=90\degree$. From Fig.~\ref{fig:lat_vec_s0p750} we can see that the lattice vectors and the angles show a sharp transition from the close-packed Minkowski lattice to the BCC lattice around $\phi \approx 0.67$. In a similar fashion we also average out the orientations and positions of individual particles in our $NPT$ simulations to construct the ideal lattice.

\subsection{\label{sec:pd}Phase Diagram}

\begin{figure*}[!htb]
\begin{center}
\includegraphics[scale=1.0]{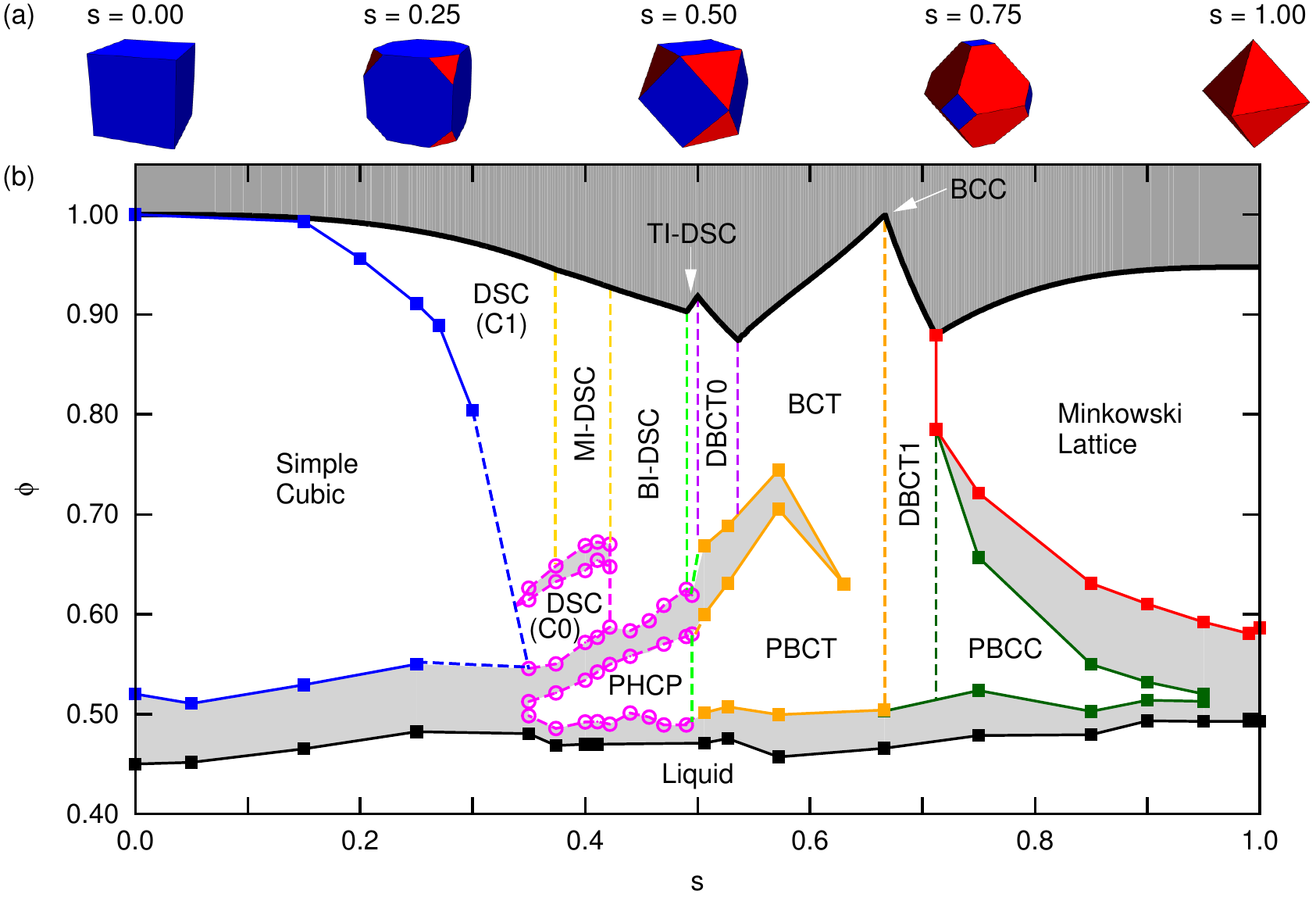}
\end{center}
\caption{\label{fig:pd} (a) Truncated cubes for five different values of the truncation parameter $s$. Truncated corners are shown in red. (b) Phase diagram for the family of truncated hard cubes in the packing fraction $\phi$ \emph{versus} shape parameter $s$ representation. In the dark-gray area $\phi$ exceeds the maximum packing fraction. The light-gray areas indicate the two-phase coexistence regions. The solid square symbols denote the bulk coexistence densities as obtained from free-energy calculations, while the open circles indicate those derived from the equations of state (EOSs). Coexistence lines that follow from free-energy calculations are represented by solid lines, and those that connect EOS derived points are given by dashed lines. The various labels stand for: distorted simple cubic (DSC), (distorted) body-centered tetragonal ((D)BCT), plastic BCT (PBCT), (plastic) body-centered cubic ((P)BCC), and plastic hexagonal close packed (PHCP). The prefixes MI-, BI-, TI- stand for mono-, bi-, and tri-interlocking, and the numbers that follow the DBCT label signify that these DBCT phases are distinct. The two DSC phases have different morphologies, one is C0-like, the other is C1-like. Finally, the two white arrows in the forbidden region connect the label TI-DSC to the small region between the green and purple dashed line and the label BCC with the turquoise line, respectively.}
\end{figure*}

As explained in the above sections, using the FBMC results in combination with regular isothermal-isobaric ($NPT$) simulations and free-energy calculations we were able to establish the full phase diagram for our hard truncated-cubes system. Figure~\ref{fig:pd} shows the phase diagram for the family of truncated cubes in the packing fraction $\phi$ \emph{vs.} the level of truncation $s$ representation. 

\begin{figure*}[!htb]
\begin{center}
\includegraphics[scale=1.]{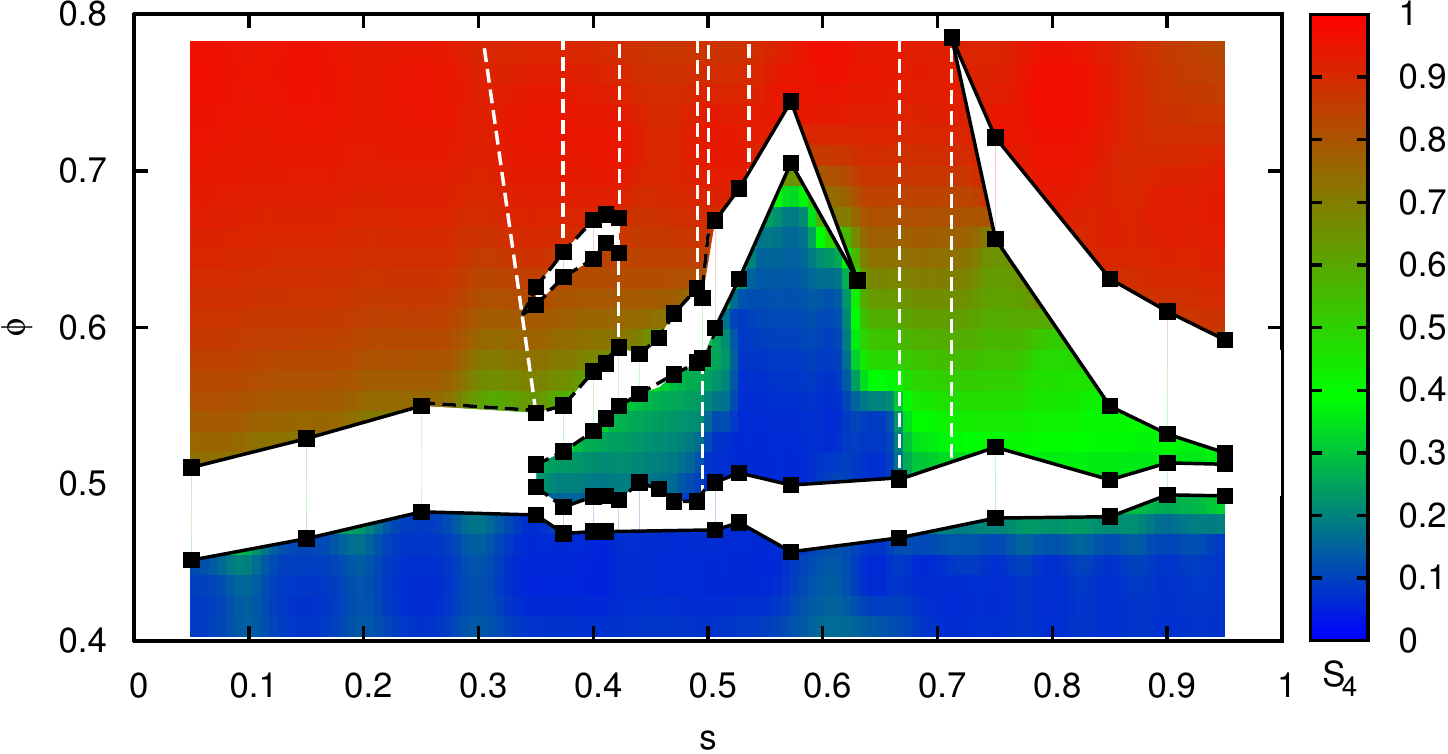}
\end{center}
\caption{\label{fig:cubatic_pd} The value of the cubatic order parameter $S_{4}$ as a function of the location in the phase diagram, where $s$ is the truncation parameter and $\phi$ is the packing fraction. The color function gives the value of $S_{4}$ as indicated by the legend next to the figure. The white regions in the plot denote the coexistence regions. The white dashed lines indicate different phase boundaries, also see Fig.~\ref{fig:pd}.}
\end{figure*}
\subsubsection{The Cubic Part}

For $s < 1/2$ the particles are essentially `cubic' in shape and we found high-density simple-cubic-like phases. The phase diagram for truncated cubes with shape parameter $s \in [0.00,0.35]$ displays three stable bulk phases. At very high pressures, we observed a distorted simple cubic (DSC) crystal phase, which is C1-like in nature, see Torquato~\emph{et al.}~\cite{Torquato_8} for the definition of the C1 structure. This phase melted either \emph{via} a weak first-order or \emph{via} a second-order phase transition into a simple cubic (SC) crystal phase. At even lower pressures, the SC crystal coexists with the fluid phase, \emph{i.e.} there is a first-order phase transition between SC and fluid. The effect of vacancies on the SC-fluid coexistence densities is not taken into account as the shift is minute. We discuss the vacancy-rich SC phase in detail in Gantapara~\emph{et al.}~\cite{Gantapara2013}

For $s\in (0.35,0.422]$ the phase diagram exhibits four stable phases, which are separated by three two-phase coexistence regions. At low pressures, we observed a liquid phase, which transformed into a plastic crystal phase with a hexagonal close-packed crystal structure (the PHCP phase) upon increasing the pressure. By further increasing the pressure the system underwent a first-order transition to a deformed simple cubic crystal (DSC) phase, which has a C0-like morphology, also see Torquato~\emph{et al.}~\cite{Torquato_8} Finally, the system self-assembled at sufficiently high pressure into the respective densest-packed structures, \emph{i.e.}, for $s\in(0.35,0.374]$ the system self-assembled into a C1-like structure (DSC) and for $s\in(0.374,0.422]$ a mono-interlocking deformed simple cubic (MI-DSC) phase is formed, as discussed in the close-packed structures. 

We found a triple point (SC/C0 $-$ PHCP $-$ liquid) at $s \approx 0.374$. For $s\in(0.422, 0.5]$ we observed higher orders of the interlocking of the DSC crystal phase at sufficiently high pressures: a bi-interlocking DSC (BI-DSC) and a tri-interlocking DSC (TI-DSC) crystal, respectively. These phases melted into the PHCP phase and subsequently into the isotropic liquid phase upon lowering the pressure, again \emph{via} first-order phase transitions in both instances.

For $s\in[0.35,0.5]$ we did not perform free-energy calculations, because there are significant fluctuations in the mean position of the particles and the averaged box vectors even for systems as large as $N\approx$ 1,000, which interfered with obtaining a proper Einstein crystal as reference system for the thermodynamic integration method,~\cite{Frenkel1984,Marechal2010} as described in the Simulation Methods section.

\subsubsection{The Octahedral Part}

For $s > 1/2$ the shape is `octahedron-like', and we found body-centered-tetragonal-like (BCT-like) structures at close packing. For $s\in[0.5,0.54]$ the close-packed distorted BCT (DBCT; labeled DBCT0, since there are multiple DBCT regions) phase melted into a plastic BCT (PBCT) phase upon lowering the pressure \emph{via} a first-order phase transition. At lower pressures, we found two-phase coexistence between the PBCT and the fluid phase. In the region $s\in(0.54,0.666]$ we obtained a regular BCT phase at high pressures, which underwent a first-order phase transition into the PBCT phase for intermediate pressures. Remarkably, for $s=2/3$ the tetragonal nature of the lattice is lost and the system exhibits a purely body-centered-cubic (BCC) crystal structure, which exists only for this exact value of the truncation parameter.

For $s\in(0.666,0.712]$ we another DBCT crystal structures (DBCT1 in Fig.~\ref{fig:pd}). All crystal structures in the region $s\in(0.636,0.712]$ melt directly into a liquid phase \emph{via} a first-order phase transition upon decreasing the pressure. That is, the coexistence region for the BCT-PBCT transition does not extend up to $s=2/3$.

In the region $s\in(0.712,0.95]$ we found a Minkowski lattice~\cite{Minkowski} at high pressures. At intermediate pressures, this system melted into a stable plastic BCC (PBCC) phase before melting into fluid. However, for $s\in(0.95,1.0]$ we found that the PBCC phase became metastable with respect to the solid-liquid phase transition (also see Ni~\emph{et al.}~\cite{Ni2012}) such that at $s=0.95$ a triple point (isotropic liquid $-$ PBCC $-$ Minkowski crystal) appeared in the phase diagram. The straight lines separating the phase boundaries for $s\in[0.374,0.712]$ at high packing fractions are a continuation of the subdivision that followed from the FBMC simulations. Several simulations close to the boundaries (on either side) are performed, to prove that within the numerical accuracy there is no deviation from the vertical phase boundaries shown in Fig.~\ref{fig:pd}.

\subsubsection{Mesophases}

\begin{figure*}[!hbt]
\begin{center}
\includegraphics[scale=1]{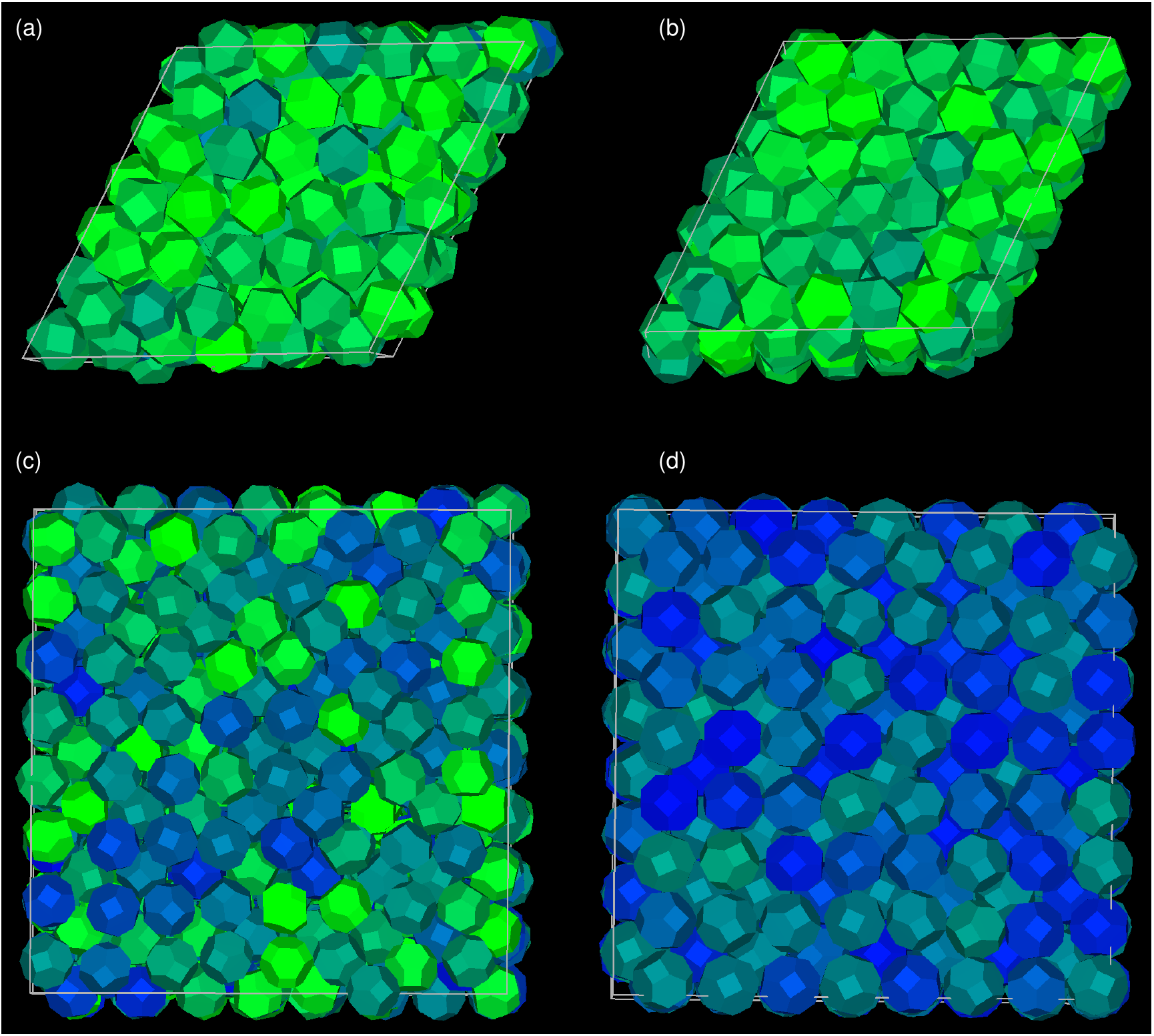}
\end{center}
\caption{\label{fig:snaps2} Snapshots of the plastic-crystal phases from our $NPT$ simulations. The plastic crystal for a truncated cube with $s=0.572$ at packing fractions close to (a) $\phi=0.545$ and (b) $\phi=0.675$ with $N=216$ particles. Typical configurations for $s=0.666$, the mathematical truncated octahedron, for packing fractions (c) $\phi=0.55$ and (d) $\phi=0.744$ with $N=1024$ particles. The coloring used here indicates the level of alignment of these particles with respect to the orientation of a reference particle in an ideal crystal. The particles which deviate maximally from the orientation of a particle in the reference ideal crystal are indicated in green and the particles with minimum deviation in blue.}
\end{figure*}

Now that we have described the position of the mesophases in the phase diagram in detail as well as the phase transitions, we will turn our attention to the order in these mesophases. We computed the cubatic order parameter $S_4$ defined in Eq.~\cref{eq:S4} as a function of packing fraction $\phi$ and shape $s$. To accomplish this we first calculated $S_4(s,\phi)$ for selected values of $s$ as a function of the pressure and in turn used this data to interpolate and determine the cubatic order $S_4(s,\phi)$ in the entire range of $s\in[0.05,0.95]$ and $\phi\in[0.4,0.8]$. We show $S_4(s,\phi)$ projected onto the phase diagram in Fig.~\ref{fig:cubatic_pd}. The use of colors is as follows: blue for $S_4(s,\phi)\approx0$, green for $S_4(s,\phi)\approx0.4$, and red for $S_4(s,\phi)\geq0.9$ and above; intermediate values are given by a smooth interpolation of these points. The white regions in Fig.~\ref{fig:cubatic_pd} represent the coexistence regions and the black squares are the exact coexistence densities calculated from the free energies. From this plot we can infer how the order develops from the freezing densities all the way up to the close-packed densities.

Figure~\ref{fig:cubatic_pd} shows that the crystal structures of truncated cubes with shape parameter $s<0.35$ develop global orientation order at relatively low packing fractions compared to the ones in the $s>0.35$ region. For truncated cubes $s\in[0.35,0.65]$ the cubatic order parameter $S_4(s,\phi)$ of the plastic crystal phases are similar to those of the isotropic fluid phase. Near $s\approx 0.58$ the cubatic order $S_4$ is less than $0.1$, even for packing fractions as high as $\phi\approx 0.69$. To give a better impression of the order in the (plastic) crystal phases, we show snapshots at packing fractions $\phi=0.545$ and $\phi=0.675$, corresponding to a plastic crystal phase both with low cubatic order for $s=0.572$ and $\phi=0.55$ and $\phi=0.744$ corresponding to BCC phase for $s=0.666$ in Fig.~\ref{fig:snaps2}.

\begin{figure*}[!hbt]
\begin{center}
\includegraphics[scale=1]{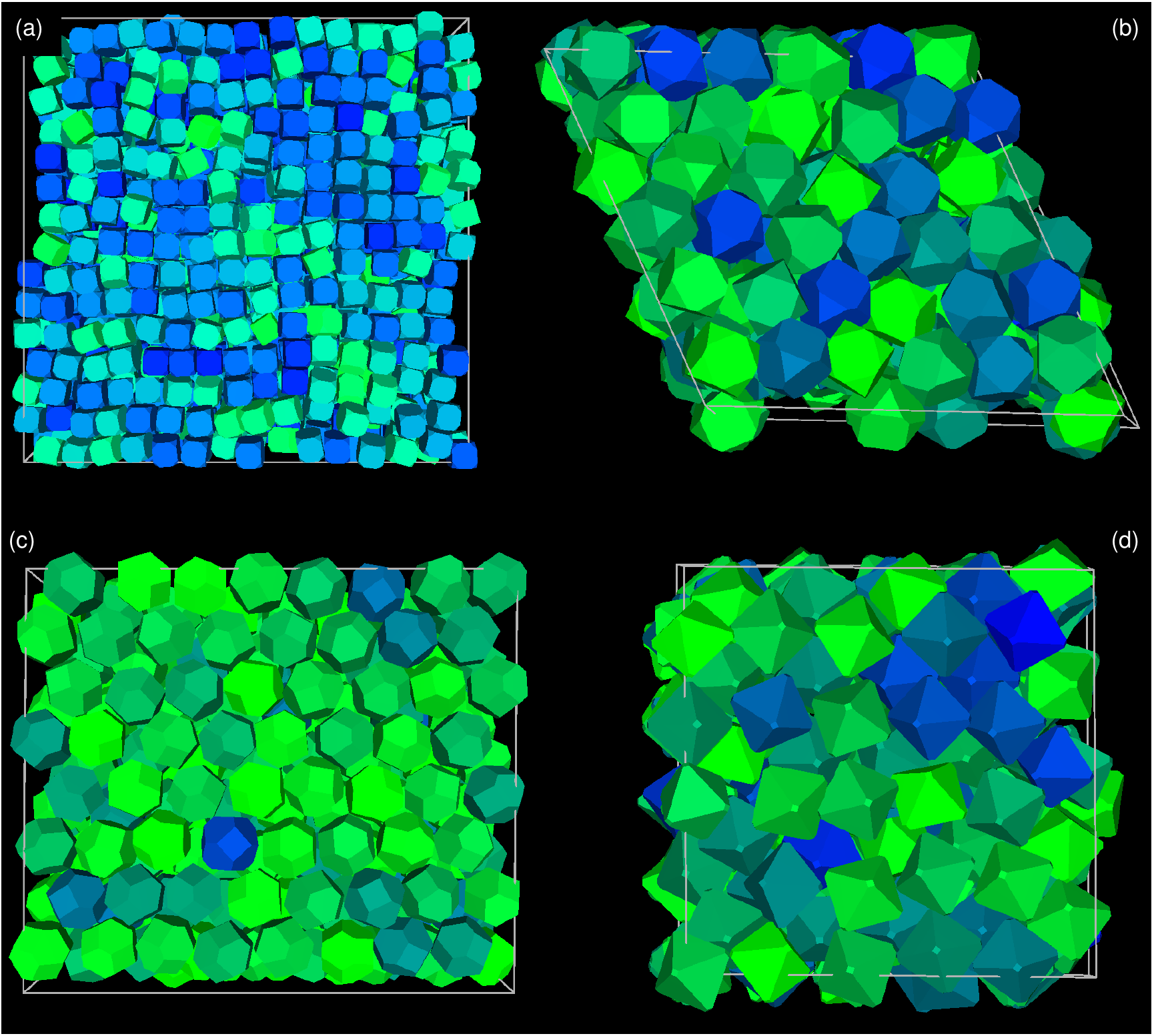}
\end{center}
\caption{\label{fig:snaps} Several snapshots of our isothermal-isobaric ($NPT$) simulations showing the various crystal structures that form in our family of truncated cubes. (a) Equilibrium vacancy-rich simple cubic crystal for a truncation of $s=0.25$ at packing fraction $\phi=0.56$. The simulation was performed for $N=$ 3,235 particles. In this system the vacancy concentration was found to be $\alpha=0.032$. (b) Plastic hexagonal-close-packed (PHCP) phase for $s=0.411$ and $\phi=0.6$ in a box containing $N=216$ particles. (c) Plastic body-centered-tetragonal (PBCT) phase for $s=0.607$ and $\phi=0.58$ in a box containing $N=512$ particles. (c) Plastic body-centered-cubic (PBCC) phase for $s=0.900$ and $\phi=0.52$ in a box containing $N=250$ particles. The coloring used here is same as the one explained in Fig.~\ref{fig:snaps2}.}
\end{figure*}

It should be pointed out that in some of the regions of Fig.~\ref{fig:cubatic_pd}, around $s\approx0.666$ and close to coexistence, the cubatic order values are as low as in a plastic-crystal phase. Our results agree with the presence of low cubatic order values for $s=0.666$ as observed by Agarwal~\emph{et al.}~\cite{Agarwal} and Thapar~\emph{et al.}~\cite{thapar2014} However, we do not consider these phases `plastic', since there is no first-order phase transition between the dense crystal and the mesophase. We show  typical configurations of a simple cubic crystal phase at $s=0.25$, a plastic HCP at $s=0.411$, a plastic BCT at $s=0.607$, and a plastic BCC phase at $s=0.900$, all slightly above fluid-solid coexistence are shown in Fig.~\ref{fig:snaps} to give (together with Fig.~\ref{fig:snaps2}) a complete impression of the mesophases that occur in the family of hard truncated cubes. 

\subsection{\label{sub:plastic_crystals}Plastic Crystal Phases}

Plastic crystals (rotator phases) are characterized by long-ranged positional order and short-ranged orientation order.~\cite{bing,timmermans,sherwood} Recent simulation studies on hard anisotropic colloidal systems have shown the existence of intriguing plastic crystalline phases.~\cite{Ni2012,Damasceno2012,Gantapara2013,Agarwal} These studies showed that the particle shape plays an important role in the formation of these plastic crystals for hard-particle systems. In addition, various physical quantities were calculated to quantify the shape of a given colloidal particle with respect to that of a sphere and to predict whether or not the particles will form a plastic crystal phase. In this section we first group the truncated cubes based on their phase behavior and their respective asphericity values. Afterwards, we describe different plastic crystals and their properties.

\begin{figure}[!htb]
\begin{center}
\includegraphics[scale=1.]{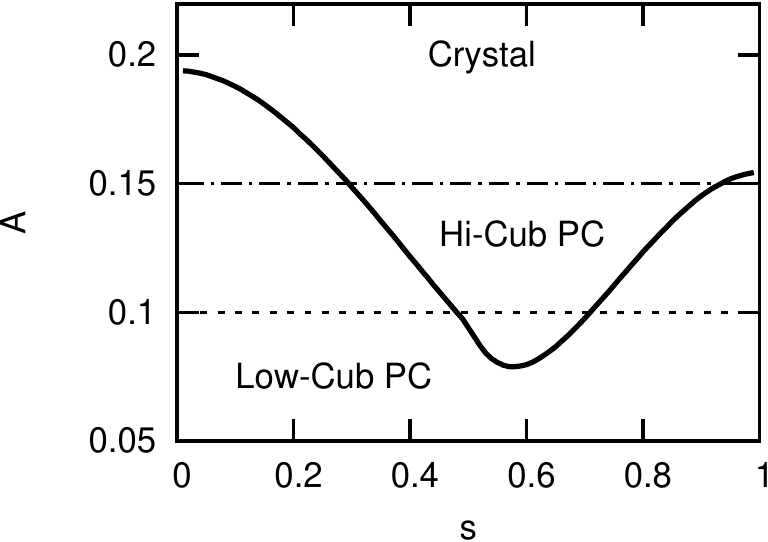}
\end{center}
\caption{\label{fig:asphericity} Asphericity $A$, defined in Eq.~\ref{eq:a}, as a function of the shape parameter $s$, which defines the level of truncation of a cube. The minimum at $s=0.58$ gives the best potential plastic-crystal former of the truncated cubes family. Based on our observations of the phase behavior of the truncated cubes we divided the $A$ into three regions Crystal, Hi-Cub~PC (plastic crystals with a high value of the cubatic order parameter), and Low-Cub~PC (plastic crystals with a low value of the cubatic order parameter). Properties of truncated cubes belonging to different regions are explained in the main text.} 
\end{figure}

Typical physical quantities used to understand the observed phase behavior of an anisotropic particle are the asphericity and the isoperimetric coefficient.~\cite{Ni2012,Agarwal,Damasceno2012} Here, we use asphericity
\begin{equation}
\label{eq:a} A = 1- \frac{\pi^{1/3}\left[6V(s)\right]^{2/3}}{S(s)}
\end{equation}
where $ V(s)$ and $S(s)$ are the volume and surface area of a truncated cube with truncation parameter $s$. The asphericity $A$ of truncated cubes as a function of the shape parameter is shown in Fig.~\ref{fig:asphericity}. 

Based on our observations of the phase behavior of truncated cubes (Fig.~\ref{fig:pd}) and the cubatic order parameter $S_4(s,\phi)$ values (Fig.~\ref{fig:cubatic_pd}) close to the fluid-crystal and fluid-plastic-crystal phase coexistence regions, we have divided the asphericity $A$ plot into three different regions. This division in terms of the asphericity is an attempt to connect the observed phase behavior to the respective asphericity values. The division is as follows.

\begin{enumerate}
\item \textbf{Crystal} Truncated cubes falling in this region freeze into a crystal phase with the cubatic order $S_4(s,\phi)\geq0.7$ when compressed from a fluid phase.
\item \textbf{Hi-Cub~PC} In this region, we observed that the truncated cubes can form plastic crystals with $S_4(s,\phi)\approx0.3-0.4$ when compressed from a fluid phase.
\item \textbf{Low-Cub~PC} The region with lowest asphericity values in the family of our truncated cubes. Truncated cubes in this region can form plastic crystals with $S_4(s,\phi)<0.1$ near the fluid-plastic-crystal phase coexistence densities.
\end{enumerate}

Using the asphericity parameter in combination with the particle's rotational symmetry one can estimate the phase behavior of anisotropic and point symmetric particles.\cite{Agarwal} The asphericity values at which we find plastic-crystal phases with low cubatic order for truncated cubes are in agreement with those of cube-like superballs, which self-assemble into plastic-crystal phases for $A<0.08$.\cite{Ni2012} We cannot make a similar comparison with octahedron-like superballs, as less is known about the phase behavior of these particles, due to the instabilities in the overlap algorithm for these superballs.\cite{Ni2012} 

\begin{figure*}[!htb]
\begin{center}
\includegraphics[scale=1]{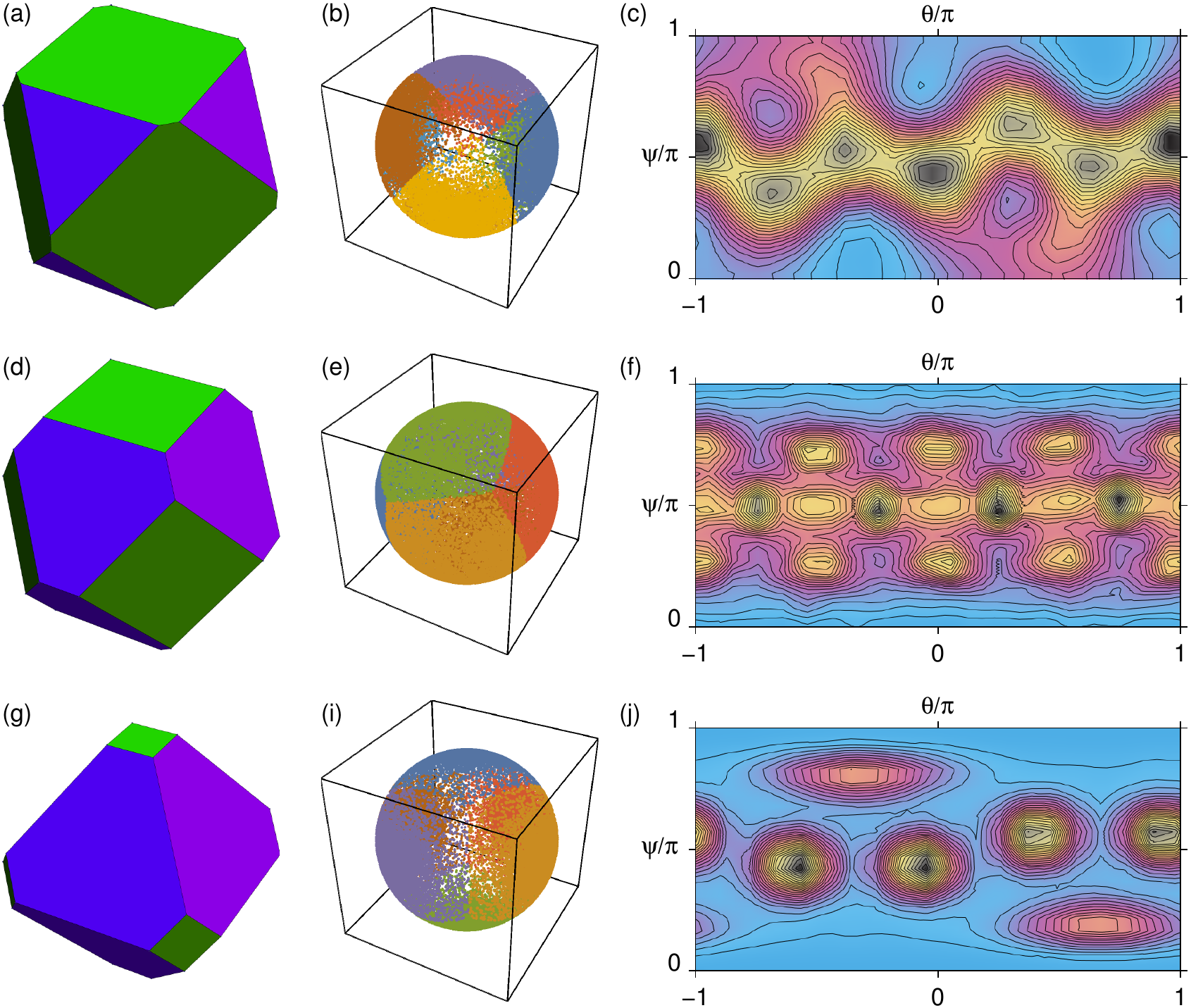}
\end{center}
\caption{\label{fig:plastic_low_phi}Various truncated cubes and their orientation distribution functions in the plastic crystal phase. Panels (a,d,g) show the particle shape for truncation $s=0.457,0.572, \text{ and } 0.750$, respectively. We have chosen these shapes to represent three different plastic crystals in the phase diagram of truncated cubes. Panels (b,e,h) show the orientations projected onto the surface of a unit sphere for the shapes shown in (a,d,g), respectively, just above the fluid-plastic crystal phase coexistence. We have colored different clusters with different (randomly chosen) colors. The clusters are obtained using the ``FindClusters'' routine of the Mathematica software package. Panels (c,f,i) show contour plots of the corresponding orientation distribution functions in the azimuthal $\theta$ and polar $\psi$ angle representation. We have used a CMYK (cyan, magenta, yellow, and black) color gradient to show the probability density of the orientation distribution functions. Low probability regions are colored cyan and high probability regions are colored black.}
\end{figure*}

In the remainder of this section we describe the properties of the plastic crystals found in the phase diagram in the region $s\in [0.35,0.95]$. We found three different types of plastic crystal, namely: HCP, BCT, and BCC. To study and understand the properties of these different plastic crystal phases we have chosen three representative particle shapes $s=0.457, 0.572, \text{ and } 0.750$. The particles with $s=0.457$ and $s=0.750$ belong to the Hi-Cub~PC region, while $s=0.572 $ lies inside the Low-Cub~PC region and is close to the minimum value of the asphericity $A$ as shown in Fig.~\ref{fig:asphericity}. The above three particle shapes are displayed in Fig.~\ref{fig:plastic_low_phi}(a,d,g), respectively.

We calculated the orientation distribution functions for these three systems at fixed pressures. Figure~\ref{fig:plastic_low_phi}(b,e,h) shows the orientation distribution function of the three particle shapes projected onto the surface of a unit sphere and Fig.~\ref{fig:plastic_low_phi}(c,f,i) shows the same distribution plotted as contour plot for our parameter choices, respectively. These orientation distribution functions are computed just above the fluid-plastic crystal coexistence region. Surprisingly, our results show that the plastic-crystal phase exhibits an inhomogeneous orientation distribution on a unit sphere. In the orientations projected onto the surface of a sphere we identified different clusters using the Mathematica ``FindClusters'' routine in combination with visual observations. These clusters are colored (randomly) to improve the clarity of the presentation, as shown in Fig.~\ref{fig:plastic_low_phi}(b,e,h). The orientation distributions show well-defined peaks for a few specific orientations dictated by the shape of the particle in combination with the crystal structure. The corresponding contour plots for the three particle shapes are shown in the azimuthal/polar ($\theta$, $\psi$) representation in Fig.~\ref{fig:plastic_low_phi}(c,f,i), respectively. The contour plots of the orientation distribution functions are colored using a CMYK color gradient, cyan (C) is used to color the low probability region, while black (K) is used to color regions with a high probability and the remaining two colors -- magenta (M) and yellow (Y) -- are used for the intermediate probabilities. Most of the peaks in the contour plots are clearly visible in Fig.~\ref{fig:plastic_low_phi}(c,f,i), however, some peaks are overshadowed by others. 

By visual inspection of the orientation distribution functions along with the results of the ``FindClusters'' routine we found that there are $6$, $16$, and $6$ distinct peaks in the orientation distribution functions for $s=0.457,0.572, \text{ and } 0.750$, respectively. Note that truncated cubes with an asphericity $A$ in the Hi-Cub~PC region ($s=0.457 \text { and } 0.750$) have a smaller number of peaks in the orientation distribution function compared to the ones with $A$ in the Low-Cub~PC region ($s=0.572$). Additionally, we found that the cubatic order is inversely proportional to the number of peaks in the orientation distribution function, \emph{i.e.}, a greater number of peaks in the orientation distribution functions gives rise to a lower cubatic order. This is due to the fact that the probability of the particles to orient themselves along one of the cubatic axes of a reference particle in the simulation box goes down if the orientation distribution function has more peaks. 

\begin{figure*}[!htb]
\begin{center}
\includegraphics[scale=1]{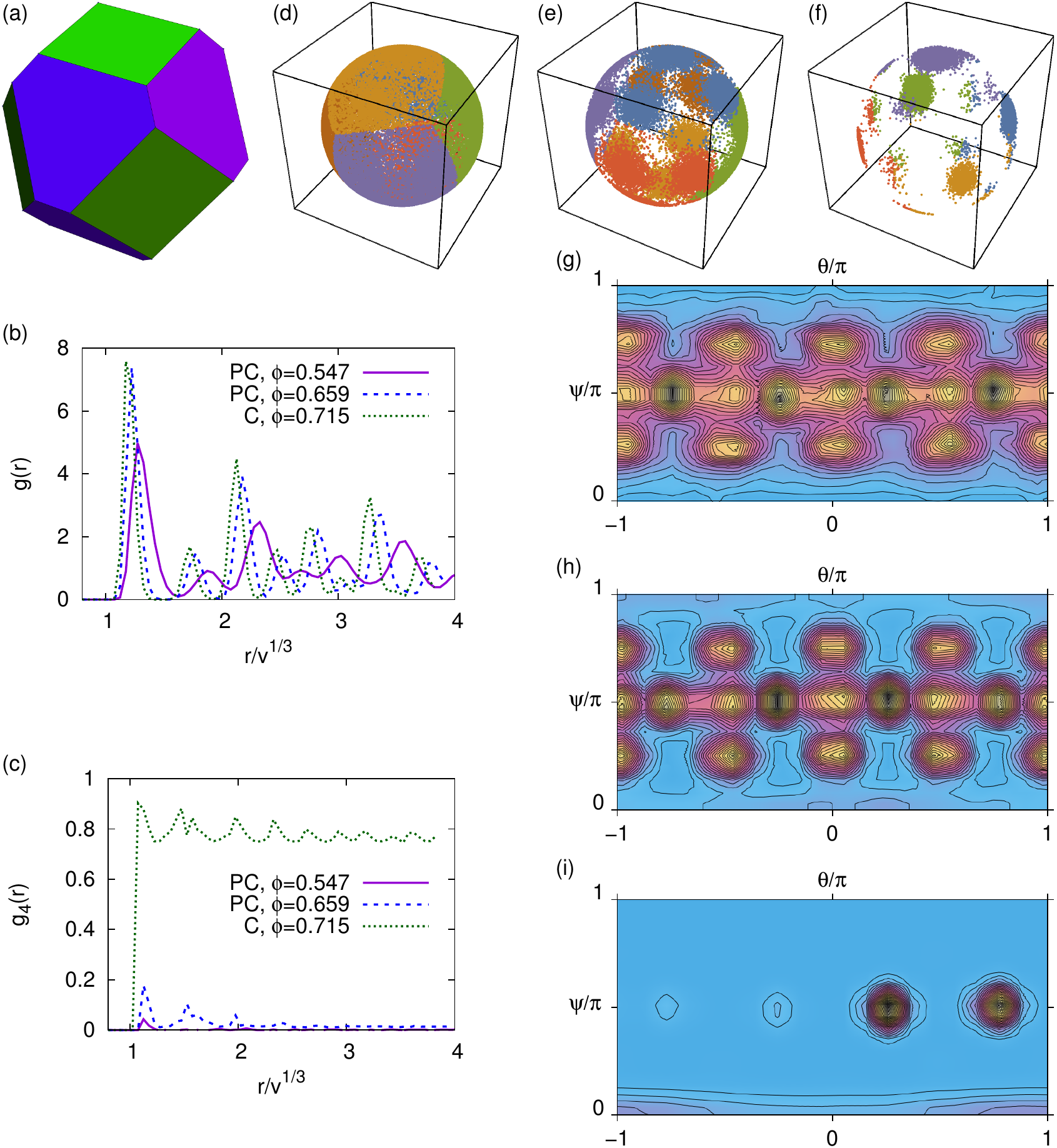}
\end{center}
\caption{\label{fig:correlations_s0p572} Analysis of particle shape $s=0.572$ as a function of packing fraction. (a) Shows the particle shape. (b) and (c) show the position and orientation correlation functions, respectively, for three different packing fractions as a function of the radial distance $r$ scaled by the cubic root of the particle volume $v$. These packing fractions are chosen such that $\phi=0.547$ is just above the liquid-plastic crystal coexistence region, while $\phi=0.659$ is slightly below the plastic-close-packed crystal coexistence region, and $\phi=0.715$ is in close-packed crystal region. In (d,e,f) we show the orientation distributions for $s=0.572$ at $\phi=0.547, 0.659, \text{ and } 0.715 $, respectively. Different clusters of orientations are (randomly) colored to improve visibility. We found that there are 18 favored orientations. Panels (g,h,i) show the density of the orientation distribution in $\theta$ and $\psi$ representation. The use of colors is the same as in Fig.~\ref{fig:plastic_low_phi}(g,h,i).} 
\end{figure*}

We found that the peaks in the orientation distribution function corresponding to the crystalline configuration increase in size with increasing packing fraction and that the peaks corresponding to non-cubatic symmetry, disappear when the system undergoes a transition from a plastic crystal to a solid phase. To further investigate this property, we study the orientation distribution of a plastic crystal as a function of packing fraction $\phi$. We chose $s=0.572$, as this particle shape has the lowest asphericity of the particles that we simulated. We show the particle shape along with its correlation function and orientation distribution functions in Fig.~\ref{fig:correlations_s0p572}. We chose three packing fractions, to calculate the correlation functions and orientation distributions: 

\begin{enumerate}
\item $\phi=0.547$, which is just above the liquid-plastic crystal coexistence region, 
\item $\phi=0.659$, which is slightly below the plastic-crystal-crystal coexistence region, 
\item $\phi=0.715$, which is in the stable crystal region. 
\end{enumerate}

In Fig.~\ref{fig:correlations_s0p572}b, we show the position correlation functions $g(r)$ for the aforementioned three packing fractions. We clearly see that $g(r)$ shows long-range positional order for all the chosen packing fractions. However, the $g_4(r)$ shown in Fig.~\ref{fig:correlations_s0p572}c exhibits long-range orientation correlations only in the crystal regime, \emph{i.e.}, for $\phi=0.715$. In the plastic-crystal phase ($\phi=0.547$ and $0.659$), the orientation correlations vanish at a distance smaller than one lattice spacing, as expected. With increasing packing fraction the orientation distribution of the particles in the plastic-crystal phase displays long-range orientation order, as shown in Fig.~\ref{fig:correlations_s0p572}. The probability density in the crystal phase ($\phi=0.715$) shows the same 16 peaks as in the plastic-crystal phase. However, the peaks close to the crystal exhibit far more sharply defined long-range orientation order.

To recap, the orientation distribution function of plastic crystals of hard anisotropic particles can be highly anisotropic and can be strongly peaked for specific orientations. These orientation directions depend not only on the crystal structure of the particle but also on the shape of the particle. Our results show that hard particle plastic crystals are different in nature from those of plastic crystals constituted of particles that have long-range interactions.~\cite{bing} Systems with long-range interactions tend to form plastic crystals with uniform orientation distribution functions unlike the hard particles studied here. 

\section{\label{sec:conclusions}Conclusions}

Summarizing, the investigation in this manuscript is a continuation of the work put forward in Ref.~\rcite{Gantapara2013}, wherein the full phase diagram was determined for a family of hard truncated cubes, which interpolates smoothly from a cube \emph{via} a cuboctahedron to an octahedron, using Monte Carlo simulations and free-energy calculations. We started our presentation by providing a detailed description of the methods employed to construct the phase diagram for our shapes and hard anisotropic faceted particles in general. Subsequently, we discussed the nature of the densest packing crystal structures, from which we later determined the equations of state (EOSs). Here, we focused on the properties of these structures, the way they can be grouped, and the differences between our grouping and that of Chen~\emph{et~al.}~\cite{chen2014} We showed that our grouping matches well with that of Chen~\emph{et~al.}, with only one minor correction to our original results. Next, the EOSs were determined by melting the densest structures and compressing from the liquid phases. Using these results in combination free-energy calculations, we established the phase diagram. This diagram shows a remarkable diversity in crystal structures. In discussing its properties, we spent special attention to the nature of the mesophases. Finally, we considered the plastic-crystalline mesophases of these hard particles in more detail and showed how their orientation distribution function display significant anisotropy. 

The following properties of the phases formed by this family of truncated cubes are of particular interest.
\begin{itemize}
\item There is a fully degenerate crystal phase for a truncation parameter $s \approx 0.4$, in which diagonally interlocked sheets of particles can move with respect to each other in only one direction. 
\item This system is remarkable in more than one way, since it also exhibits a fluid state and three different bulk crystals upon increasing the pressure. Both these qualities may make similarly shaped nanoparticles suitable for the creation of highly tunable functional materials, for which optical, electrical, and rheological properties vary strongly with the bulk pressure of the system. 
\item We calculated the cubatic order parameter $S_4$ for truncated cubes with varying truncation level and showed that the values of $S_4$ are related to the number of preferred particle orientations in the plastic-crystal phase. We found that PHCP and PBCT plastic-crystal phases in the truncated-cube family have $S_4$ values similar to that of the isotropic fluid and exhibit short range orientation correlations for asphericity values $A<0.1$. This is a surprising result for faceted particles, as this behavior was expected only for particles with smooth edges.~\cite{Ni2012}
\item A comparison to the results for superballs~\cite{Ni2012,Torquato_8} leads us to conclude that plastic crystal (or rotator) phases of faceted particles have a smaller domain of stability. Moreover, the phase behavior as a function of shape parameter is much smoother for hard superballs than for truncated cubes. These observations give rise to the idea that the more spread-out local curvature of the superball tends to favor the formation of rotator phases and overall smoother phase behavior, whereas the polyhedral particles with flat faces and sharp edges prefer to align flat faces to form crystals and have sharp transitions even though $s$ varies smoothly. 
\item Our study shows that the orientation distribution for particles in plastic-crystal phases can be highly anisotropic. This anisotropic nature of the orientation distribution for hard particle plastic crystals shows that these phases are fundamentally different in nature from those of plastic crystals constituted by particles that have long-range interactions, as the latter exhibit a homogeneous orientation distribution. The favored orientations depend not only on the particle shape, but also on the crystal structure. In addition, we found that the cubatic order of the plastic-crystal phases is inversely proportional to the number of peaks in the orientation distribution functions. 
\end{itemize}

Our results provide a solid basis for future studies of anisotropic particle systems and pave the way for a full understanding of the recent experimental studies performed on systems of nanoscopic truncated cubes. In addition, our study of the phase behavior of truncated cubes by smoothly varying the shape can be used in future studies to obtain rules for the prediction of self-assembled structures based only on the shape.

\bibliography{thesis}
\end{document}